\documentclass[journal,10pt,a4paper]{IEEEtran}
\usepackage{graphicx}
\usepackage{amsmath}
\usepackage{amssymb}
\usepackage{amsfonts}
\usepackage{bm}
\usepackage{upgreek}
\usepackage{multirow}
\usepackage{xcolor}
\usepackage[normalem]{ulem}

\newcommand{\bd}{\mathbf}

\newcommand{\be}{_{\mathrm e}}
\newcommand{\bt}{_{\mathrm t}}
\newcommand{\us}{^{\mathrm s}}

\newcommand{\txs}{_{\mathrm{t_x}}^{\mathrm s}}
\newcommand{\tys}{_{\mathrm{t_y}}^{\mathrm s}}
\newcommand{\tzs}{_{\mathrm{t_z}}^{\mathrm s}}

\newcommand{\exs}{_{\mathrm{e_x}}^{\mathrm s}}
\newcommand{\eys}{_{\mathrm{e_y}}^{\mathrm s}}
\newcommand{\ezs}{_{\mathrm{e_z}}^{\mathrm s}}

\newcommand{\ps}{_{\mathrm p}^{\mathrm s}}

\newcommand{\ts}{_{\mathrm t}^{\mathrm s}}

\newcommand{\es}{_{\mathrm e}^{\mathrm s}}

\newcommand{\lp}{_{\mathrm {perm}}^{\mathrm s}}
\newcommand{\li}{_{\mathrm {ind}}^{\mathrm s}}
\newcommand{\lE}{_{\mathrm{eddy}}^{\mathrm s}}

\newcommand{\bTL}{_{\mathrm{TL}}}
\newcommand{\bimp}{_{\mathrm{AUG}}}
\newcommand{\bKFF}{_{\mathrm{KFF}}}
\newcommand{\bs}{_{\mathrm{s}}}

\IEEEoverridecommandlockouts
\begin{document}

\title{Fusion-based Dynamic Platform Magnetic Compensation: Beyond the Tolles–Lawson Approach}
\author{Rong Yang\\
        Yaakov Bar-Shalom\\ 
\thanks{Authors’ addresses: R. Yang, DSO National Laboratories, 12~Science Park Drive, Singapore~118225 (E-mail: yrong@dso.org.sg); Y. Bar-Shalom, Department of ECE, University of Connecticut, Storrs, CT 06269, USA (E-mail:yaakov.bar-shalom@uconn.edu).}
}
\markboth{ }%
{Shell \MakeLowercase{\textit{et al.}}: Bare Demo of IEEEtran.cls for Journals}

\maketitle

\begin{abstract}
 Aeromagnetic compensation for airborne platforms is essential for real-time tracking of the dynamic external magnetic field free from platform interference, enabling robust magnetic navigation (MagNav), magnetic anomaly detection (MAD), and other geophysics applications. While the classical Tolles-Lawson (TL) framework focuses extensively on estimating compensation model parameters, it pays less attention to the real-time estimation of the dynamic external magnetic field. To bridge this gap, this paper proposes a two-stage calibration and compensation framework. First, an augmented linearized model is developed for sensor-based calibration parameter estimation, avoiding the information loss inherent in classical band-pass filtering (BPF). Second, utilizing these pre-estimated parameters, a map-less Kalman Filter Fusion (KFF) algorithm is developed to dynamically estimate the external field and compensate the platform interference directly from multi-sensor measurements corrupted by platform interference. The second step enables real-time joint dynamic estimation and multi-sensor compensation without requiring reference position data (which runs counter to the MagNav purpose) or prior anomaly maps. Validated on the public DAF-MIT MagNav dataset, the framework overcomes the non-causal limitations of existing approaches, reducing average compensation error from 359.2~nT to 12.3~nT while maintaining exceptional robustness against heavily corrupted sensor channels.
\end{abstract}

\section{Introduction}\label{s1}

Magnetic field measurements are essential for airborne geophysical surveying, magnetic anomaly navigation (MagNav), and magnetic anomaly detection (MAD). In practical airborne systems, however, magnetometer measurements are corrupted by platform-generated magnetic interference arising from ferromagnetic materials, onboard electronics, electrical currents, and aircraft maneuvers. The removal of such interference, commonly referred to as aeromagnetic compensation, remains a fundamental challenge in airborne magnetic sensing.

The foundation of modern aeromagnetic compensation traces back to the pioneering work of Tolles and Lawson \cite{Tolles1950}, who developed one of the first practical compensation schemes for MAD-equipped aircraft (with a detailed derivation provided in \cite{Gnadt2022}). The classical TL model represents aircraft magnetic interference through permanent, induced, and eddy-current field components, typically formulated as an 18-parameter model. By applying linearization, physical approximations, and band-pass filtering (BPF), the TL methodology enables calibration parameters to be estimated efficiently via a linear Least Squares (LS) estimator. Owing to its simplicity, physical interpretability, and practical effectiveness, the TL approach remains the dominant methodology for airborne magnetic compensation.

The TL estimation problem, however, is generally ill-conditioned. In practical evaluations, the condition number of the estimation matrix can reach up to $10^7$, even when executing recommended calibration flight paths \cite{Gnadt2022}. Leliak \cite{Leliak1961} demonstrated that the original 18-parameter TL model can be reduced to 16 independent parameters, presenting a more compact representation to mitigate parameter redundancy. Since then, both 18-parameter and 16-parameter models have been widely used in the literature. In this work, the 18-parameter model is selected, as experimental demonstrations indicate that both models yield comparable condition numbers, while the 18-parameter model offers superior physical interpretability.

To improve compensation robustness, several researchers have revisited the modeling of the external magnetic field within the classical TL framework. Rather than eliminating the external field purely through BPF and approximations, recent studies advocate incorporating the International Geomagnetic Reference Field (IGRF)~\cite{Alken2021} alongside Inertial Navigation System (INS) attitude tracking to model the external field in the platform body frame~\cite{Ye2024, Liu2025}. This mapping eliminates the need for a physical vector magnetometer and are claimed to outperform traditional fluxgate-based mapping. However, constructing this IGRF+INS mapping introduces three critical, compounding error sources:
\begin{enumerate}
    \item \textbf{Magnitude Discrepancy:} The IGRF value deviates significantly from the true local external field because regional crustal anomalies and diurnal variations are omitted. The norm difference between external and IGRF fields ($| \bd B\be | - | \bd B_{\text{IGRF}} |$) can easily exceed $100\,\text{nT}$, even during high-altitude calibration flights at $3\,\text{km}$.
    \item \textbf{INS Attitude Errors:} Roll, pitch, and yaw tracking uncertainties introduce orientation-dependent transformation errors.
    \item \textbf{Angular Mismatch:} Small angular deviations ($\delta \theta$) persist between the theoretical IGRF vector and the true external field vector $\bd B\be$. As demonstrated in a recent study~\cite{Hager2026sub}, this directional mismatch alone ($\approx B_e \cdot \delta \theta$) can induce over 87~nT of projection error.
\end{enumerate}
In this work, we still select direct vector magnetometer mapping to prevent these synthetic projection errors from degrading compensation performance.

Moreover, calibration parameters estimated offline from batch measurements often suffer performance degradation when platform internal environmental conditions drift or abruptly change. This discrepancy has driven significant interest in online calibration methodologies. Recursive Least Squares (RLS) has been applied directly to online variants of the TL model~\cite{Dou2018}. Online parameter tracking was subsequently integrated within a navigation filter~\cite{Bonifaz2020}, with real-time magnetic navigation later demonstrated on an F-16 platform~\cite{Canciani2022}. These online calibration concepts have also been extended to machine learning frameworks~\cite{Hager2026}. However, many of these approaches rely on map-aiding or platform position feedback, which runs counter to the primary purpose of standalone MagNav.

Furthermore, the limitations of purely physics-based compensation models have motivated various machine-learning approaches, including neural networks \cite{Hezel2020, Jiang2025, Hager2026}, Random Forests \cite{Moradi2024}, and Temporal Convolutional Networks \cite{Wang2025}. Although these data-driven methods capture complex nonlinearities, they require substantial training data and lack physical interpretability. More fundamentally, because many machine-learning approaches couple with the TL model to form Physics-Informed Neural Networks (PINNs) \cite{Gnadt2022t}, their performance remains bounded by the inherent approximations of the underlying physical formulation. This highlights that fundamental research in baseline physical modeling remains crucial to unlocking the full potential of both classic and learning-augmented approaches.

Concurrently, research initiatives at the MIT Artificial Intelligence Accelerator (AIA) and the Air Force Institute of Technology (AFIT) have significantly advanced the field through the rigorous development of the classical TL model and machine learning approaches~\cite{Gnadt2022t}, alongside the public release of benchmark resources—specifically the DAF-MIT MagNav dataset and open-source software packages~\cite{Gnadt2023, MagNavjl2022}.

Other advancements include vector magnetic navigation~\cite{Canciani2020}, addressing vector magnetometer calibration errors~\cite{Han2017} and magnetic gradient effects~\cite{Ge2021, Feng2022}.

Despite these advances, existing model-based approaches remain primarily focused on estimating calibration parameters and removing platform interference from individual sensors, devoting comparatively little attention to the real-time, dynamic estimation of the external magnetic field itself. Furthermore, existing methodologies generally perform compensation independently for each sensor and fail to explicitly exploit multi-sensor fusion.

To address these limitations, a two-stage magnetic compensation framework is proposed in this paper. First, the calibration parameter estimation is reformulated through an augmented, map-based parameter estimation model. This model overcomes the structural limitations inherent in the classical TL formulation, yielding accurate, sensor-specific parameters without relying on conventional simplifications and band-pass filtering. Second, leveraging these calibrated parameters, a multi-sensor Kalman filter fusion (KFF) scheme is developed. By formulating the external magnetic field as a dynamic state, the KFF scheme enables the joint real-time estimation of the external field and multi-sensor compensation. This KFF process is entirely map-less, requiring no prior reference data from the IGRF, magnetic anomaly maps, or diurnal base stations. To the best of our knowledge, such an approach has not been previously reported in the aeromagnetic literature.

The main contributions of this paper are summarized as follows:
\begin{enumerate}
    \item \textbf{Augmented Linear Calibration Model:} An augmented linear model is introduced to reformulate the offline parameter estimation problem by incorporating explicit sensor measurement biases and replacing the total field proxy with external field projections. This eliminates structural information loss inherent in conventional band-pass filtering.
    
    \item \textbf{Map-Less Multi-Sensor KFF Compensation:} A real-time, map-less KFF compensation scheme is proposed to dynamically track the external magnetic field while simultaneously compensating for platform interference across multiple sensors. This eliminates the need for prior magnetic anomaly maps, IGRF projections, or external position aiding.

   \item \textbf{Deep Analysis of Ill-Conditioned Calibration and Robust Compensation:} Rigorous evaluation on the DAF-MIT MagNav flight dataset reveals why parameter sets from ill-conditioned calibration problems still yield highly accurate downstream compensation. Additionally, empirical results highlight the necessity of online calibration to account for temporal parameter variations. Furthermore, detailed compensation analysis proves the exceptional robustness of the KFF scheme against heavily corrupted sensor channels.

    \item \textbf{Exploration of Alternative Mathematical Approaches:} Two alternative approaches—Iterated Least Squares (ILS) applied to the nonlinear calibration model and a vector-based linear calibration model—are developed. Although these formulations appear conceptually superior and more advanced, empirical analysis on real flight data reveals that they fail to match the accuracy of the proposed linear model, providing key insights into the practical limitations of these formulations.
\end{enumerate}

The remainder of this paper is organized as follows. Section II reviews the classical TL formulation. Section III analyzes its fundamental limitations. Section IV presents the proposed calibration parameter estimation approach. Section V develops the multi-sensor KFF algorithm for real-time dynamic external field estimation. Section VI evaluates the proposed framework using real flight dataset. Section VII discusses alternative approaches investigated during this research, and Section VIII concludes the paper.

\section{Tolles--Lawson Approach}\label{s2}

First, the primary mathematical notations used throughout this paper are summarized in Table~\ref{tb_1}.

\begin{table}[!htbp]
\caption{Notations}
\label{tb_1}
\centering
\begin{tabular}{l p{6.2cm}}
\hline\hline
\textbf{Symbol} & \textbf{Description} \\
\hline
$\bd B\ts$ & Total magnetic field vector at sensor position in the sensor frame; $\bd B\ts=[B\txs, B\tys, B\tzs]'$, where $(\cdot)'$ denotes transposition. \\
$\bd u\ts$ & Unit vector along the direction of $\bd B\ts$. \\
$B\bt$     & Magnitude of total magnetic field, $B\bt = \|\bd B\ts\|$. \\
$\bd B\es$ & External magnetic field vector in sensor frame, $\bd B\es=[B\exs, B\eys, B\ezs]'$. \\
$\bd B\be$ & External magnetic field vector in the North-East-Down (NED) frame. \\
$B\be$     & Magnitude of external field, $B\be = \|\bd B\be\| = \|\bd B\es\|$. \\
$\bd B\ps$ & Platform-induced magnetic field vector in sensor frame. \\
$\bd B\lp$ & Permanent magnetic field component of $\bd B\ps$. \\
$\bd B\li$ & Induced magnetic field component of $\bd B\ps$. \\
$\bd B\lE$ & Eddy-current magnetic field component of $\bd B\ps$. \\
\hline\hline
\end{tabular}
\end{table}

The TL model is the classical linear framework for airborne magnetic compensation, widely employed to remove aircraft-induced magnetic interference. It typically pairs a high-precision scalar magnetometer with a tri-axial vector magnetometer to estimate and remove maneuver-dependent disturbances from the scalar measurements. Owing to its computational simplicity and practical efficacy, the TL model remains a standard benchmark in aeromagnetic surveying. The mathematical derivation and structure of the classical TL model are reviewed below.

\begin{figure}[!htbp]
	\centering
	\includegraphics[width=2.0in]{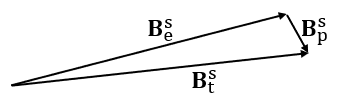}
	\caption{Vector relationship between the external magnetic field $\bd B\es$, platform-induced interference $\bd B\ps$, and the total measured magnetic field $\bd B\ts$ in the sensor coordinate frame.}
	\label{f1}
\end{figure}

The fundamental vector magnetic field model is illustrated in Fig.~\ref{f1} and expressed as
\begin{equation} \label{eq_tol}
    \bd{B}\ts = \bd{B}\es + \bd{B}\ps
\end{equation}
where all quantities are expressed in the vector magnetometer frame. These field vectors vary continuously over time; however, the explicit discrete-time index $k$ is temporarily omitted here for brevity. The time index will be formally reintroduced later in this section to formulate the batch parameter estimation in (\ref{eq_TL}).

The platform-induced magnetic field $\bd B\ps$ is decomposed into permanent, induced, and eddy-current components
\begin{align}\label{eq_Ba}
    \bd B\ps &= \bd{B}\lp + \bd{B}\li + \bd{B}\lE \nonumber \\
             &= \bd a + \bd b \bd{B}\es + \bd c \dot{\bd{B}}\es
\end{align}
where
\begin{equation}
\bd{a} = 
\begin{bmatrix} 
a_x \\ 
a_y \\ 
a_z 
\end{bmatrix}\ \ 
\bd{b} = 
\begin{bmatrix} 
b_{xx} & b_{xy} & b_{xz} \\ 
b_{xy} & b_{yy} & b_{yz} \\ 
b_{xz} & b_{yz} & b_{zz} 
\end{bmatrix}\ \ 
\bd{c} = 
\begin{bmatrix} 
c_{xx} & c_{xy} & c_{xz} \\ 
c_{yx} & c_{yy} & c_{yz} \\ 
c_{zx} & c_{zy} & c_{zz} 
\end{bmatrix}
\end{equation}
Here, $\bd a \in \mathbb{R}^3$ represents the permanent hard-iron vector, $\bd b \in \mathbb{R}^{3 \times 3}$ is a symmetric matrix modeling soft-iron induced magnetization, and $\bd c \in \mathbb{R}^{3 \times 3}$ models eddy-current effects.

To obtain a scalar calibration model, the magnitude relationship is linearized using the law of cosines:
\begin{align}
\|\bd{B}\es\|
&= \sqrt{\|\bd{B}\ts\|^2 - 2\,\bd{B}\ts \cdot \bd{B}\ps + \|\bd{B}\ps\|^2} \nonumber \\
&= \|\bd B\ts\| \sqrt{1-\frac{2\,\bd{B}\ts \cdot \bd{B}\ps}{\|\bd{B}\ts\|^2} + \frac{\|\bd{B}\ps\|^2}{\|\bd{B}\ts\|^2}} \nonumber \\
&\approx \|\bd B\ts\| \sqrt{1-\frac{2\,\bd{B}\ts \cdot \bd{B}\ps}{\|\bd{B}\ts\|^2}}
\end{align}
Applying a first-order Taylor series expansion yields for the magnitudes (which do not need coordinate system superscript)
\begin{equation}\label{eq_linear}
B\bt \approx B\be + (\bd{u}\ts)' \bd{B}\ps
\end{equation}
where
\begin{equation}\label{eq_Bhat}
  \bd u\ts = [u\txs, u\tys, u\tzs]' = \left[\frac{B\txs}{B\ts}, \frac{B\tys}{B\ts}, \frac{B\tzs}{B\ts}\right]'
\end{equation}
denotes the unit vector along the total magnetic field direction, expressed in the vector magnetometer coordinate frame.

A critical classical approximation then replaces $\bd B\es$ in (\ref{eq_Ba}) with $\bd B\ts$:
\begin{equation}\label{eq_Ba_app}
    \bd B\ps \approx \bd a + \bd b \bd{B}\ts + \bd c \dot{\bd{B}}\ts
\end{equation}
where the unknown external field $\bd{B}\es$ is approximated by the total measured field $\bd{B}\ts$ as a proxy. Substituting (\ref{eq_Bhat}) and (\ref{eq_Ba_app}) into (\ref{eq_linear}) yields the explicit permanent, induced, and eddy-current terms detailed below.

\subsubsection*{Permanent term}

\begin{equation}
(\bd u\ts)' \, \bd{a}
= a_x u\txs + a_y u\tys + a_z u\tzs
\end{equation}

\subsubsection*{Induced term}

\begin{align}
(\bd u\ts)'\,\bd{b}\,\bd{B}\ts
&=
\begin{bmatrix}
u\txs & u\tys & u\tzs
\end{bmatrix}
\bd{b}
\begin{bmatrix}
B\txs \\ B\tys \\ B\tzs
\end{bmatrix} \nonumber \\
&=b_{xx} u\txs B\txs  + b_{yy}  u\tys B\tys + b_{zz}  u\tzs B\tzs\nonumber \\
&+ 2b_{xy} u\txs B\tys
+ 2b_{xz} u\txs B\tzs
+ 2b_{yz} u\tys B\tzs
\end{align}

\subsubsection*{Eddy-current term}

\begin{equation}
(\bd u\ts)'\,\bd{c}\,\dot{\bd{B}}\ts
=
\sum_{i,j \in \{x,y,z\}} c_{ij}\,u_i\us \,\dot{B}_j\us
\end{equation}
Stacking over time from time $1$ to $n$, yields the TL linear model
\begin{equation}\label{eq_TL}
\bd y = \bd y\be + \mathbf{A}\bTL\bd{\uptheta}\bTL +\bd V
\end{equation}
where the parameter vector to be estimated is
\begin{align}\label{eq_TLpara}
   \bd{\uptheta}\bTL =& \left[a_x\ a_y\ a_z\ b_{xx}\ b_{yy}\ b_{zz} \ b_{xy}\ b_{xz}\ b_{yz}\right.\nonumber\\
   &\left. c_{xx}\ c_{xy}\ c_{xz}\ c_{yx}\ c_{yy}\ c_{yz}\ c_{zx}\ c_{zy}\ c_{zz}\right]'
\end{align}
and $\bd V$ is the batch measurement zero-mean Gaussian noise.
The batch of scalar magnetometer measurements and external fields are, respectively,
\begin{align}
\mathbf{y} &=
\begin{bmatrix}
B\bt(1) & \cdots & B\bt(n)
\end{bmatrix}'\\
\mathbf{y}_e &=
\begin{bmatrix}
B\be(1) & \cdots & B\be(n)
\end{bmatrix}'\label{eq_BeStuck}
\end{align}
The matrix $\mathbf{A}\bTL$ is given by
\begin{align}\label{eq_TL_A}
\mathbf{A}\bTL &=
\begin{bmatrix}
\bd A\bTL(1) \\
\vdots \\
\bd A\bTL(n)
\end{bmatrix}
\end{align}
with
\begin{align}\label{eq_TLAk}
   \bd A\bTL(k) = [&u\txs(k)\ \ u\tys(k)\ \ u\tzs(k)\nonumber\\
   &u\txs(k) B\txs(k)\ \ u\tys(k) B\tys(k)\ \ u\tzs(k) B\tzs(k)\nonumber\\
   &2u\txs(k) B\tys(k)\ \ 2u\txs(k) B\tzs(k)\ \ 2u\tys(k) B\tzs(k)\nonumber\\
   &u\txs(k) \dot B\txs(k)\ \ u\txs(k) \dot B\tys(k)\ \ u\txs(k)\dot B\tzs(k)\nonumber \\
   &u\tys(k) \dot B\txs(k)\ \ u\tys(k) \dot B\tys(k)\ \ u\tys(k)\dot B\tzs(k)\nonumber \\
   &u\tzs(k) \dot B\txs(k)\ \ u\tzs(k) \dot B\tys(k)\ \ u\tzs(k)\dot B\tzs(k)]
\end{align}
where $k$ is the time index; $B\txs(k)$, $B\tys(k)$, and $B\tzs(k)$ are the orthogonal components measured by the vector magnetometer at time $k$; and $u\txs(k)$, $u\tys(k)$, and $u\tzs(k)$ are the corresponding unit vector components.

To estimate $\bd{\uptheta}\bTL$ using the TL model in~(\ref{eq_TL}), the external field component $\bd y_e$ remains unknown. A bpf is therefore applied to remove both low- and high-frequency components, under the assumption that $\bd y_e$ is concentrated in the low-frequency band, while measurement noise in both scalar and vector magnetometers is mainly high-frequency. This leads to
\begin{align}
   \mathrm{bpf}(\bd y - \bd y_e) &= \mathrm{bpf}(\bd A\bTL\bd{\uptheta}\bTL+\bd V),\\
   \mathrm{bpf}(\bd y) &= \mathrm{bpf}(\bd A\bTL)\bd{\uptheta}\bTL
\end{align}
The parameter $\bd{\uptheta}\bTL$ can be estimated using the regularized LS method
\begin{equation}
    \hat{\bd{\uptheta}}\bTL = \left[\mathrm{bpf}(\bd A\bTL)' \mathrm{bpf}(\bd A\bTL) + \lambda \mathbf{I}\right]^{-1} \mathrm{bpf}(\bd A\bTL)' \mathrm{bpf}(\bd y)
\end{equation}
The regularization term is included because the problem is typically ill-conditioned. The regularized LS estimator is more numerically stable than the standard LS estimator.

\section{Limitations of the Tolles–Lawson Approach}\label{s3}

Although the TL model has served as the foundational framework for aircraft magnetic compensation for several decades, a close inspection of its derivation reveals several inherent limitations. Furthermore, to evaluate these classical assumptions against real-world conditions, the DAF-MIT MagNav dataset \cite{Gnadt2023} is utilized as an empirical baseline. An evaluation of the multi-sensor cabin magnetometer data against high-fidelity reference readings reveals a distinct mismatch between the theoretical assumptions of the TL model and actual aircraft interference fields. This section details these specific mathematical and physical limitations.

\subsection{Lack of Real-Time Compensation Capability}
First, the platform compensation evaluation step itself is not rigorously addressed within the classical TL framework. For instance, in the open-source software implementation MagNav.jl~\cite{MagNavjl2022}, compensation is computed as
\begin{equation}\label{eq_TL_eval}
\bd y\be = \bd y - \bd A\bTL\bd{\uptheta}\bTL + \overline{\bd A\bTL \bd{\uptheta}\bTL}
\end{equation}
which attempts to recover low-frequency signal components attenuated during band-pass filtering by adding the batch mean term $\overline{\bd A\bTL \bd{\uptheta}\bTL}$. However, this heuristic approach is physically inconsistent and unsuitable for real-time applications, as evaluating the mean across an entire flight segment introduces non-causal dependencies that rely on future temporal data unavailable during online operation.

\subsection{Structural Mismatch and Information Loss from Band-Pass Filtering}
Second, the band-pass filter (bpf) removes both low- and high-frequency components of the measurements. While this suppresses low frequency spatial external field variations and high-frequency sensor noise, it inadvertently removes physically meaningful low-frequency components of the platform interference field---specifically those contributed by the permanent field $\bd{B}\lp$ and the induced field $\bd{B}\li$. Consequently, filtering alters the underlying structural properties of $\bd{B}\ps$, yet the unfiltered model \eqref{eq_Ba} is still directly applied to the filtered data for parameter estimation, introducing a fundamental structural mismatch.

To evaluate whether the bpf can effectively isolate the platform field from external field variations, a numerical simulation was conducted using the trajectory from flight \texttt{1002.02} of the MIT MagNav dataset~\cite{Gnadt2023}, representing a calibration flight at an altitude of $3\text{~km}$. The International Geomagnetic Reference Field (IGRF) is assumed as the ground-truth external field and is computed along the flight path using latitude, longitude, altitude, and timestamp data. This IGRF vector is transformed into the platform coordinate frame using attitude angles (roll, pitch, yaw) from the inertial navigation system (INS). The platform interference field is then generated via \eqref{eq_Ba} with parameters $\bd{a} = [500, 10, 200]'$, the elements of $\bd{b}$ ranging between $-0.001$ and $0.01$, and the elements of $\bd{c}$ randomly sampled from $\mathcal{N}(0, 10^{-10})$.

\begin{figure}[!htbp]
	\centering
	\includegraphics[width=3.25in]{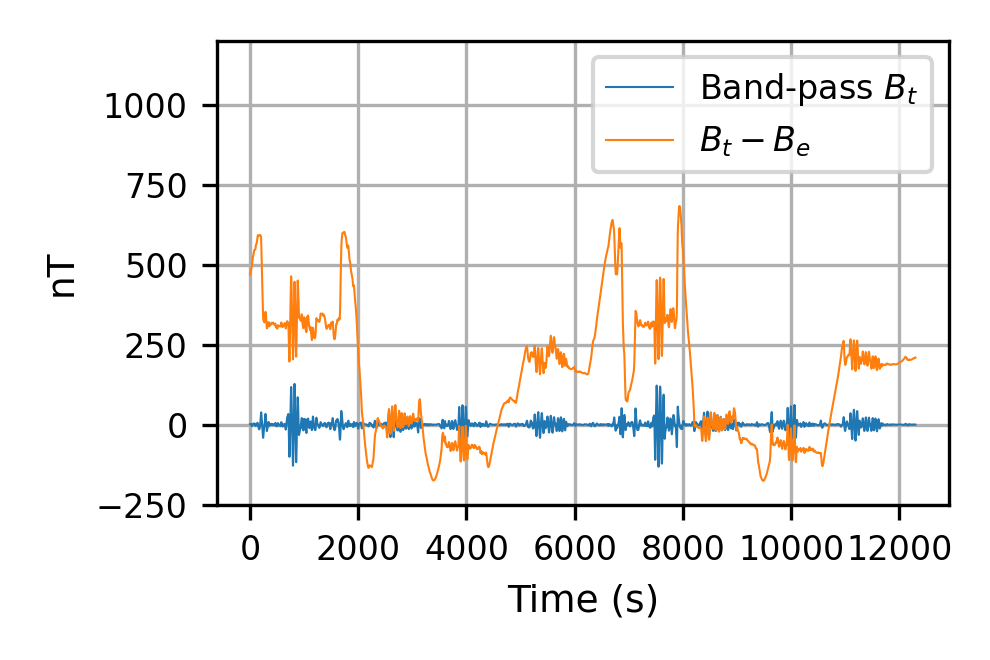}
	\caption{Comparison between the band-pass-filtered total magnetic field $B\bt$ and the actual platform interference signal $B\bt - B\be$.}
	\label{f2}
\end{figure}

The simulation results are presented in Fig.~\ref{f2}. As illustrated, the band-pass-filtered total field variation (blue curve) severely deviates from the true target signal $B\bt - B\be$ (orange curve), confirming that critical low-frequency platform interference is erroneously suppressed by the filtering operation.

\subsection{Modeling Error from External Field Proxy Substitution}
Third, the classical TL framework approximates the unknown external magnetic field $\bd B\es$ using the total measured field $\bd B\ts$ (which incorporates both the external field $\bd B\es$ and the platform interference field $\bd B\ps$) in \eqref{eq_Ba_app}. This proxy substitution introduces a non-negligible modeling error. Because the platform interference model itself depends directly on the vector relationship and magnitude difference between $\bd B\es$ and $\bd B\ts$ (as illustrated in Fig.~\ref{f1}), substituting the corrupted measurement $\bd B\ts$ in place of the true driving field $\bd B\es$ biases the estimated aircraft interference parameters.

\subsection{Spatial Non-Colocation of Vector and Scalar Magnetometers}
Fourth, the classical TL framework estimates the calibration parameters of the scalar magnetometer using directional measurements provided by the vector magnetometer. In practical airborne platforms, these two sensors are rarely co-located. Consequently, they experience distinct localized magnetic field environments, rendering their respective measurements inconsistent with a single, shared set of calibration parameters. The classical assumption that scalar and vector measurements correspond to a single spatial point introduces structural inconsistency when jointly utilizing both sensors for parameter estimation.

\subsection{Omission of Explicit Magnetometer Bias Terms}
Fifth, the magnetometers exhibit distinct sensor measurement biases across different flights in the MIT MagNav dataset \cite{Gnadt2023}. Although these static bias offsets are effectively suppressed during band-pass filtering, failing to model them explicitly leaves the estimation framework incomplete. Incorporating a dedicated bias state directly within the platform field model allows the sensor bias to be jointly estimated alongside the aircraft interference parameters. Explicitly recovering and removing this bias is essential for achieving optimal, unbiased platform compensation when processing raw, unfiltered measurements in real-time.

\vspace{1em}
It is worth noting that prior studies \cite{Feng2022, Ge2021} often assume that the TL model requires a constant external magnetic field $\bd B\es$. While this assumption is necessary for certain extended formulations involving $\bd B\es$, the derivation presented in Section~\ref{s2} imposes no such restriction. In this work, $\dot{\bd B}\es$ inherently accounts for time-varying field fluctuations arising from platform maneuvering, spatial geomagnetic variations (e.g., IGRF and local anomalies), and temporal field dynamics, which can be directly evaluated via numerical differentiation.

\section{Magnetometer Calibration Parameter Estimation}\label{s4} 

The calibration parameter vector, as defined in~(\ref{eq_TLpara}), depends heavily on the sensor's physical location and the local magnetic environment of the platform. Consequently, each scalar or vector magnetometer requires a separate calibration parameter vector to be estimated. Through an empirical analysis of the DAF-MIT MagNav dataset, we observed that magnetometer measurements typically exhibit an unknown, persistent bias that remains unmodeled by the classical TL formulation. To account for this discrepancy, we introduce an explicit bias term $\beta$ and define the augmented calibration parameter vector as
\begin{align}\label{eq_B1para} 
\bd{\uptheta}\bimp =& \left[\beta\ a_x\ a_y\ a_z\ b_{xx}\ b_{yy}\ b_{zz}\ b_{xy}\ b_{xz}\ b_{yz}\right.\nonumber\\ &\left. c_{xx}\ c_{xy}\ c_{xz}\ c_{yx}\ c_{yy}\ c_{yz}\ c_{zx}\ c_{zy}\ c_{zz}\right]' \end{align} 

Accordingly, the scalar magnetometer measurement model is formulated as
\begin{align}\label{eq_modelScalar}
z(k) =& B\bt(k) + \beta + v(k)\nonumber\\ 
=& \beta + |\bd B\es(k) + \bd a + \bd b \bd B\es(k) + \bd c \dot{\bd B}\es(k)| + v(k)
\end{align} 
where $k$ denotes the discrete-time index, $z(k)$ is the scalar sensor measurement as time $k$, and $v(k) \sim \mathcal{N}(\bd{0},R)$ represents a zero-mean white Gaussian measurement noise. 

For a vector magnetometer, the measurement model is expressed as
\begin{equation}\label{eq_modelV}
    \bd z(k) = \beta \bd u\ts(k) + \bd B\es(k) + \bd a + \bd b\,\bd B\es(k) + \bd c\,\dot{\bd B}\es(k) + \bd v(k)
\end{equation}
where $\bd z(k)$ and $\bd v(k)$ denote the vector magnetometer measurement and its corresponding measurement error vector, respectively.

To estimate these calibration parameters, the external magnetic field vector $\bd B\es(k)$ must be determined. We assume that the external magnetic field intensity, $B\be(k)$, is known \emph{a priori} by aggregating the IGRF, the diurnal field variation, and the local magnetic anomaly map contribution. This assumption holds provided that no dynamic magnetic sources exist along the calibration trajectory. 

Two approaches can be employed to resolve the vector $\bd B\es(k)$. The first approach extracts the unit vector $\bd u\ts(k)$ directly from the raw vector magnetometer measurement $\bd z(k)$, yielding the approximation
\begin{equation}\label{eq_approxBe}
\bd B\es(k) \approx B\be(k)\,\bd u\ts(k)
\end{equation} 
This formulation is valid because the platform's disturbing fields are orders of magnitude smaller than the Earth's dominant background field, rendering the directional deviation between $\bd B\ts(k)$ and $\bd B\es(k)$ negligible.

The second approach transforms the reference IGRF unit vector into the platform's body coordinate system using Inertial Navigation System (INS) attitude estimates (roll, pitch, and yaw), and subsequently maps it to the sensor coordinate system via a fixed transformation to align it with $\bd z(k)$. Evaluations using DAF-MIT MagNav data showed that the first approach is more accurate. The second approach is primarily limited by attitude uncertainties inherent in the INS estimates. Therefore, the first approach is adopted for the remainder of this paper.

The scalar model in~(\ref{eq_modelScalar}) is nonlinear and can be solved for $\bd{\uptheta}\bimp$ using an Iterated Least Squares (ILS) estimator~\cite{BarShalom2001} over a batch of measurements collected from time $1$ to $n$. The vector model in~(\ref{eq_modelV}) is linear, allowing its stacked batch formulation to be solved directly via a standard LS estimator. However, both approaches yield inaccurate results, which will be demonstrated in Section~\ref{s6}. The scalar-based estimation formulation constitutes an ill-conditioned estimation
problem; because ILS requires multiple matrix inversion operations, it is prone to numerical instability and yields poor estimation performance. The vector model suffers from a higher matrix condition number, which degrades parameter observability. Furthermore, compounded by the larger 3D measurement errors inherent to vector magnetometers, parameter estimation accuracy is significantly degraded.

To address these limitations, we linearize the scalar measurement model. Following a derivation similar to the conventional TL formulation, applying the law of cosines together with a first-order Taylor series expansion yields
\begin{align}\label{eq_Abik}
    z(k) \approx &  B\be(k) + \beta + \bd u\ts(k)'\bd B\es(k) + v(k)\nonumber\\
    = & B\be(k) + \beta + \bd u\ts(k)'[\bd a + \bd b \bd B\es(k) + \bd c \dot{\bd B}\es(k)] + v(k)\nonumber\\
    = & B\be(k) +\bd A\bimp(k) \bd{\uptheta}\bimp + v(k)
\end{align}
with the augmented ``measurement" matrix 
\begin{align}\label{eq_B1Ak}
   \bd A\bimp(k) = [&1\ \ u\txs(k)\ \ u\tys(k)\ \ u\tzs(k)\nonumber\\
   &u\txs(k) B\exs(k)\ u\tys(k) B\eys(k)\ \ u\tzs(k) B\ezs(k)\nonumber\\
   &2u\txs(k) B\eys(k)\ \ 2u\txs(k) B\ezs(k)\ \ 2u\tys(k) B\ezs(k)\nonumber\\
   &u\txs(k) \dot B\exs(k)\ \ u\txs(k) \dot B\eys(k)\ \ u\txs(k)\dot B\ezs(k)\nonumber \\
   &u\tys(k) \dot B\exs(k)\ \ u\tys(k) \dot B\eys(k)\ \ u\tys(k)\dot B\ezs(k)\nonumber \\
   &u\tzs(k) \dot B\exs(k)\ \ u\tzs(k) \dot B\eys(k)\ \ u\tzs(k) \dot B\ezs(k)]
\end{align}
Compared to $\bd A\bTL(k)$ in~(\ref{eq_TLAk}), $\bd A\bimp(k)$ appends a $1$ as its first element for the bias term $\beta$. Additionally, the external field components $B\exs(k)$, $B\eys(k)$, and $B\ezs(k)$ in $\bd A\bimp(k)$ replace the total field components $B\txs(k)$, $B\tys(k)$, and $B\tzs(k)$ in $\bd A\bTL(k)$.

Stacking the measurements from time $1$ to $n$ yields the batch form
\begin{equation}
    \bd Z = \bd y\be + \bd A\bimp \bd{\uptheta}\bimp + \bd V
\end{equation}
where $\bd y\be$ is defined in~(\ref{eq_BeStuck}), and the batch measurement vector, mesurement noise and regressor matrix are
\begin{align}
\bd Z =
\begin{bmatrix}
z(1) & \cdots & z(n)
\end{bmatrix}'
\end{align}
\begin{align}
\bd V =
\begin{bmatrix}
v(1) & \cdots & v(n)
\end{bmatrix}'
\end{align}
\begin{align}
\mathbf{A}\bimp &=
\begin{bmatrix}
\bd A\bimp(1) \\
\vdots \\
\bd A\bimp(n)
\end{bmatrix}
\end{align}

This linear batch model can be solved directly using the standard Least LS estimator as
\begin{equation}\label{eq_LS1}
    \hat{\bd{\uptheta}}\bimp = \left(\mathbf{A}\bimp' \mathbf{A}\bimp\right)^{-1} \mathbf{A}\bimp' \left(\bd Z - \bd y\be\right)
\end{equation}

However, if the regressor matrix $\mathbf{A}\bimp$ is ill-conditioned due to insufficient trajectory dynamics or collinearity in the features, the matrix inversion becomes numerically unstable. To mitigate this, a regularized least squares solution (Tikhonov regularization) can be employed:
\begin{equation}\label{eq_LS2}
    \hat{\bd{\uptheta}}\bimp = \left(\mathbf{A}\bimp' \mathbf{A}\bimp + \lambda \mathbf{I}\right)^{-1} \mathbf{A}\bimp' \left(\bd Z - \bd y\be\right)
\end{equation}
where $\lambda > 0$ is the regularization parameter and $\mathbf{I}$ represents the identity matrix of appropriate dimensions.

For a vector magnetometer, this same linearized scalar estimation framework can be applied to its total field scalar intensity $B\bt$, expressed as
\begin{equation}
    B\bt = \sqrt{(B\txs)^2+(B\tys)^2+(B\tzs)^2}
\end{equation}
As demonstrated in Section~\ref{s6}, this yields significantly higher calibration parameter estimation accuracy than using the vector model presented in~(\ref{eq_modelV}).

\section{External magnetic field dynamic estimation with Fusion}\label{s5}

This section develops a dynamic estimation algorithm to compute the external magnetic field magnitude $B\be$ in real time. The approach, designated as Kalman Filter Fusion (KFF), leverages a standard Kalman filter to fuse multi-sensor measurements. It assumes that the calibration parameters for each scalar or vector magnetometer have been estimated \emph{a priori} using the methodology presented in Section~\ref{s4}. The proposed method requires at least one vector magnetometer; in multi-magnetometer scenarios, measurements from all sensors are fused to jointly contribute to the state estimate. Notably, prior knowledge of the IGRF background field and local magnetic anomaly maps is not required.

The state vector is defined as
\begin{equation}
\bd{x}(k)=[ B\be(k)\ \ \dot{B}\be(k)]'
\end{equation}
where $k$ is the discrete-time index. A nearly constant-rate dynamic model\footnote{This is the magnetic counterpart of the nearly constant velocity motion model used extensively in estimation of dynamic system. The driving white noise makes the state a Markov process, a necessary and sufficient condition to use recursive estimation \cite{BarShalom2001, BarShalom2011}.} is assumed:
\begin{equation}
\bd{x}(k) = \bd{F}\bd{x}(k-1) + \bd{w}(k-1)
\end{equation}
where
\begin{equation}
\bd{F} =
\begin{bmatrix}
1 & \Delta t\\
0 & 1
\end{bmatrix}
\end{equation}
$\Delta t$ is the sampling interval, and $\bd{w}(k) \sim \mathcal{N}(\bd{0},\bd{Q})$ represents the zero-mean white Gaussian process noise with covariance $\bd Q=$diag$[Q_{11},Q_{22}]$ with units nT$^2$ and (nT/s)$^2$.

For a configuration comprising $n\bs$ magnetometers, the measurement vector, which carries out the fusion, is given by
\begin{equation}\label{eq_Z}
\bd{z}(k) = [z_1(k)\ \ \ldots\ \ z_{n\bs}(k)]'
\end{equation}
where $z_i(k)$ for $i\in\{1,\ldots,n\bs\}$ is the $i$th magnetometer measurement at time $k$. If the $i$th sensor is a vector magnetometer, $z_i(k)$ is taken as the Euclidean norm of the measured vector. Adapting the linearized formulation~(\ref{eq_Abik}) established in Section~\ref{s4}, the global measurement model is structured as
\begin{equation}
\bd{z}(k) = \bd{H}(k)\bd{x}(k) + \bd B(k) + \bd{v}(k)
\end{equation}
Correspondingly, the individual measurement equation for the $i$th sensor is given by
\begin{equation}\label{eq_z1}
z_i(k) = \bd{H}_i(k)\bd{x}(k) + B_i(k) + v_i(k)
\end{equation}
where $v_i(k) \sim \mathcal{N}(0,R_i)$ is the white Gaussian measurement noise with variance $R_i$. The explicit derivations for the time-varying measurement row vector $\bd H_i(k)$ and the offset scalar $B_i(k)$ are detailed next.

This approach leverages the body-frame unit vector $\bd u\ts(k)$ measured by a vector magnetometer to project the scalar external field states back into their respective vector components within the sensor frame, thereby establishing the structural requirement for at least one vector sensor configuration. The mapping is expressed as:
\begin{align}
    \bd B\es(k) &= B\be(k)\bd u\ts(k) \label{eq_map1}\\
    \dot{\bd B}\es(k) &= \dot B\be(k) \bd u\ts(k) + B\be(k) \dot{\bd u}\ts(k) \label{eq_map2}
\end{align}
where the time derivative $\dot{\bd u}\ts(k)$ can be obtained via numerical differentiation and is structured as the vector:
\begin{equation}
    \dot{\bd u}\ts(k) = [\dot u\txs(k) \quad \dot u\tys(k) \quad \dot u\tzs(k)]'
\end{equation}

Substituting the mapping relations~(\ref{eq_map1}) and~(\ref{eq_map2}) into the linearized parameter matrix expression~(\ref{eq_Abik}) isolates the dynamic tracking states $B\be(k)$ and $\dot B\be(k)$. Rearranging terms yields the linear measurement formulation
\begin{align}
z_i(k) =& B\be(k) + \beta_i + \bd u\ts(k)'\left[\bd a_i + \bd b_i B\be(k)\bd u\ts(k)\right.\nonumber\\
&\left. + \bd c_i \dot B\be(k) \bd u\ts(k) + \bd c_i B\be(k) \dot{\bd u}\ts(k)\right]+ v_i(k)\nonumber\\
=& [1+\bd u\ts(k)' \bd b_i \bd u\ts(k) +\bd u\ts(k)' \bd c_i \dot{\bd u}\ts(k)]B\be(k)\nonumber\\
& + \bd u\ts(k)' \bd c_i \bd u\ts(k)\dot B\be(k)\nonumber\\
& + \beta_i + \bd u\ts(k)' \bd a_i + v_i(k)
\end{align}
Comparing this formulation directly with~(\ref{eq_z1}) yields:
\begin{align}
\bd{H}_i(k) &= \begin{bmatrix} 
1 + \bd u\ts(k)' \bd b_i \bd u\ts(k) + \bd u\ts(k)' \bd c_i \dot{\bd u}\ts(k) \\ 
\bd u\ts(k)' \bd c_i \bd u\ts(k) 
\end{bmatrix}' \\
B_i(k) &= \beta_i + \bd u\ts(k)' \bd a_i
\end{align}

With the analytical forms of $\bd{H}_i(k)$ and $B_i(k)$ resolved, the measurement model in~(\ref{eq_Z}) is fully defined by stacking the individual rows and scalars for all $n\bs$ sensors (to carry out their fusion):
\begin{align}
\bd{H}(k) &= \begin{bmatrix} \bd{H}_1(k) \\ \vdots \\ \bd{H}_{n\bs}(k) \end{bmatrix}, \quad \bd B(k) = \begin{bmatrix} B_1(k) \\ \vdots \\ B_{n\bs}(k) \end{bmatrix}
\end{align}

[[start delete here]]

The dynamic state estimation problem is thus fully formulated with strictly linear state transition and measurement models. Consequently, optimal real-time estimation of the state vector $\bd{x}(k)$ can be executed recursively using the standard Kalman filter (KF):

\subsubsection*{Prediction Step}
\begin{align}
    \hat{\bd{x}}(k|k-1) &= \bd{F}\hat{\bd{x}}(k-1|k-1) \label{eq_state_pred} \\
    \bd{P}(k|k-1) &= \bd{F}\bd{P}(k-1|k-1)\bd{F}^T + \bd{Q} \label{eq_cov_pred}
\end{align}

\subsubsection*{Update Step}
\begin{align}
    \tilde{\bd{z}}(k) &= \bd{z}(k) - \bd{H}(k)\hat{\bd{x}}(k|k-1) - \bd{B}(k) \label{eq_innovation} \\
    \bd{S}(k) &= \bd{H}(k)\bd{P}(k|k-1)\bd{H}^T(k) + \bd{R} \label{eq_innov_cov} \\
    \bd{K}(k) &= \bd{P}(k|k-1)\bd{H}^T(k)\bd{S}^{-1}(k) \label{eq_kalman_gain} \\
    \hat{\bd{x}}(k|k) &= \hat{\bd{x}}(k|k-1) + \bd{K}(k)\tilde{\bd{z}}(k) \label{eq_state_update} \\
    \bd{P}(k|k) &= \left[\bd{I} - \bd{K}(k)\bd{H}(k)\right]\bd{P}(k|k-1) \label{eq_cov_update}
\end{align}

It should be noted that potential residual errors in the \emph{a priori} parameter estimate $\hat{\bd{\uptheta}}\bimp$, alongside measurement noise in the directional cosines $\bd{u}\ts(k)$ and their time derivatives $\dot{\bd{u}}\ts(k)$, are not explicitly modeled in $\bd Q$ and $\bd R$. The current linear formulation assumes these unmodeled error sources exert a minor impact on estimation accuracy; if empirical validation indicates non-negligible performance degradation, the perfect solution exceeds the standard KF, and new algorithm needs to be developed.

[[end delete here]]

The dynamic estimation problem is thus fully formulated, where both the state transition and measurement models are strictly linear. Consequently, the optimal real-time estimation of the state can be executed directly using the standard Kalman Filter (KF), designated as KF with Fusion (KFF), detailed below.

Prediction:
\begin{align}
    \hat{\bd{x}}(k|k-1) &= \bd{F}\hat{\bd{x}}(k-1|k-1) \label{eq_state_pred} \\
    \bd{P}(k|k-1) &= \bd{F}\bd{P}(k-1|k-1)\bd{F}' + \bd{Q} \label{eq_cov_pred}
\end{align}

Update:
\begin{align}
    \tilde{\bd{z}}(k) &= \bd{z}(k) - \bd{H}(k)\hat{\bd{x}}(k|k-1) - \bd{B}(k) \label{eq_innovation} \\
    \bd{S}(k) &= \bd{H}(k)\bd{P}(k|k-1)\bd{H}^T(k) + \bd{R} \label{eq_innov_cov} \\
    \bd{K}(k) &= \bd{P}(k|k-1)\bd{H}(k)'\bd{S}^{-1}(k) \label{eq_kalman_gain} \\
    \hat{\bd{x}}(k|k) &= \hat{\bd{x}}(k|k-1) + \bd{K}(k)\tilde{\bd{z}}(k) \label{eq_state_update} \\
    \bd{P}(k|k) &= \left[\bd{I} - \bd{K}(k)\bd{H}(k)\right]\bd{P}(k|k-1) \label{eq_cov_update}
\end{align}

It should be noted that potential errors in the \emph{a priori} parameter estimate $\hat{\bd{\uptheta}}\bimp$, alongside measurement noise in the directional cosines $\bd{u}\ts(k)$ and their time derivatives $\dot{\bd{u}}\ts(k)$, are not explicitly modeled in the noise covariances $\bd Q$ and $\bd R$. The current linear formulation assumes these unmodeled error sources have a minor impact on estimation accuracy. If empirical validation indicates non-negligible performance degradation, these unmodeled error sources will be addressed as a topic for future study.

\section{Experimental Results on Real Data}\label{s6}

This section evaluates the performance of the proposed algorithms using the DAF-MIT MagNav dataset (Flight 1002). The experimental validation is conducted in two phases: first, the calibration parameters are estimated using the linearized method derived in Section~\ref{s4}; second, these estimated parameter vectors are utilized to track the external magnetic field~$B\be$ based on the dynamic state estimation model proposed in Section~\ref{s5}.

\subsection{Experimental Setup and Dataset Description}

To evaluate the performance of the proposed algorithms, experimental validation is conducted using real flight data from the publicly available DAF-MIT MagNav dataset. Specifically, data from Flight 1002, collected via a Cessna Grand Caravan aircraft over the Ottawa region, is selected for analysis. This flight path (trajectory) consists of multiple distinct legs characterized by varying flight profiles: two dedicated, highly maneuvering calibration legs executed at high altitude (\texttt{1002.02} and \texttt{1002.20}), and multiple low-altitude legs. 

The aircraft instrumentation includes a comprehensive suite of five scalar magnetometers (\texttt{mag\_1} through \texttt{mag\_5}) and four three-axis vector fluxgate magnetometers (\texttt{flux\_a} through \texttt{flux\_d}). The scalar magnetometer \texttt{mag\_1} is mounted on the tail stinger of the aircraft, and its professionally compensated output, denoted as \texttt{mag\_1\_c}, provides an accurate external magnetic field intensity. This signal serves as the definitive reference ground truth for evaluating the external field estimation performance of $B\be$. Due to data collection anomalies, \texttt{mag\_2} and \texttt{flux\_a} did not operate properly during this flight and are excluded from our analysis.

\begin{figure}[ht]
	\centering
	\includegraphics[width=2.5in]{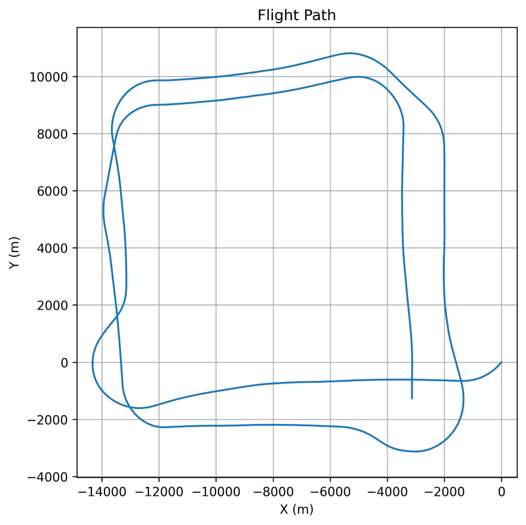}
	\caption{Two-dimensional horizontal projection of flight path for calibration \texttt{1002.02}, originating at $(0,0)$ at a nominal altitude of 3000~m. The path forms a closed box-like calibration loop designed to execute highly dynamic pitch, roll, and yaw maneuvers along its edges.}
	\label{f3}
\end{figure}

In our experiments, trajectory \texttt{1002.02} is selected as the calibration trajectory to estimate the parameter vector~$\bd{\uptheta}\bimp$ for each operational magnetometer. As illustrated by the X-Y projection in Fig.~\ref{f3}, this flight trajectory forms a closed box-like calibration loop at a nominal altitude of 3000~m, where each straight edge incorporates highly dynamic rotational excitation across the pitch, roll, and yaw angles to isolate the aircraft's internal magnetic interference.

To evaluate the performance of the calibration parameter estimation and dynamic external field intensity state estimation, 14 distinct flight trajectories are selected from the remaining data (\texttt{1002.03} through \texttt{1002.11}, \texttt{1002.13}, \texttt{1002.15}, \texttt{1002.18}, \texttt{1002.19}, and \texttt{1002.20}). Utilizing these diverse navigation trajectories provides a robust validation across varying flight dynamics and environmental conditions.

\subsection{Calibration parameter Estimation Test}\label{s6a}

The estimation of the calibration parameters based on~(\ref{eq_LS1}) or~(\ref{eq_LS2}) requires constructing the regressor matrix $\bd{A}\bimp$, which depends on the sensor-frame orientation unit vector $\bd{u}\ts$ derived from a vector magnetometer. The \texttt{flux\_b} sensor is utilized in this study, though alternative fluxgate sensors may be deployed without a significant difference in performance. Another critical component required for estimation is the true external field magnitude $B_{\mathrm{e}}(k)$. In operational environments, this is typically obtained by combining a regional magnetic anomaly map, the IGRF, and diurnal variation data. However, because a spatial anomaly map was unavailable for this specific evaluation dataset, the professionally compensated tail-stinger measurement \texttt{mag\_1\_c} is directly utilized as the external field reference during the offline parameter estimation phase.

During testing, the estimation problem was found to be severely ill-conditioned, with the condition number of the regressor matrix $\bd{A}\bimp$ reaching approximately $10^7$. This high collinearity leads to extreme numerical instability and overfitted solutions in standard LS estimation. To mitigate this ill-conditioning and enforce numerical stability, the regularized LS framework formulated in~(\ref{eq_LS2}) is used instead of the standard version~(\ref{eq_LS1}). The regularization parameter was empirically selected as $\lambda = 0.1$ to balance residual minimization with parameter bounds.

Directly evaluating the accuracy of the estimated parameter vector $\hat{\bd{\uptheta}}\bimp$ is impossible since its true physical values are unknown. Therefore, an indirect validation method is adopted: the parameter accuracy is evaluated via the Root Mean Square Error (RMSE) of the compensated residuals as
\begin{equation}
    \text{RMSE}(\hat{\bd{\uptheta}}\bimp) = \sqrt{\frac{1}{n} \sum_{k=1}^{n} \left[ \hat{B}\be(k)\bimp - B\be(k) \right]^2}
\end{equation}
where $k$ is the discrete time index, $n$ is the total number of measurements for a given sensor on the flight trajectory, and $B\be(k)$ is the corresponding reference ground-truth value obtained from \texttt{mag\_1\_c}. The compensated external field magnitude $\hat{B}\be(k)\bimp$ is computed as
\begin{equation}\label{eq_Bimp}
    \hat{B}\be(k)\bimp = z(k) - \bd{A}\bimp(k)\hat{\bd{\uptheta}}\bimp
\end{equation}
where $z(k)$ is the raw sensor measurement. If $\hat{\bd{\uptheta}}\bimp$ successfully captures the aircraft interference, the compensated profiles will consistently align with the reference baseline, even when the platform undergoes intense maneuvering.

\begin{figure}[ht]
	\centering
	\includegraphics[width=3.3in]{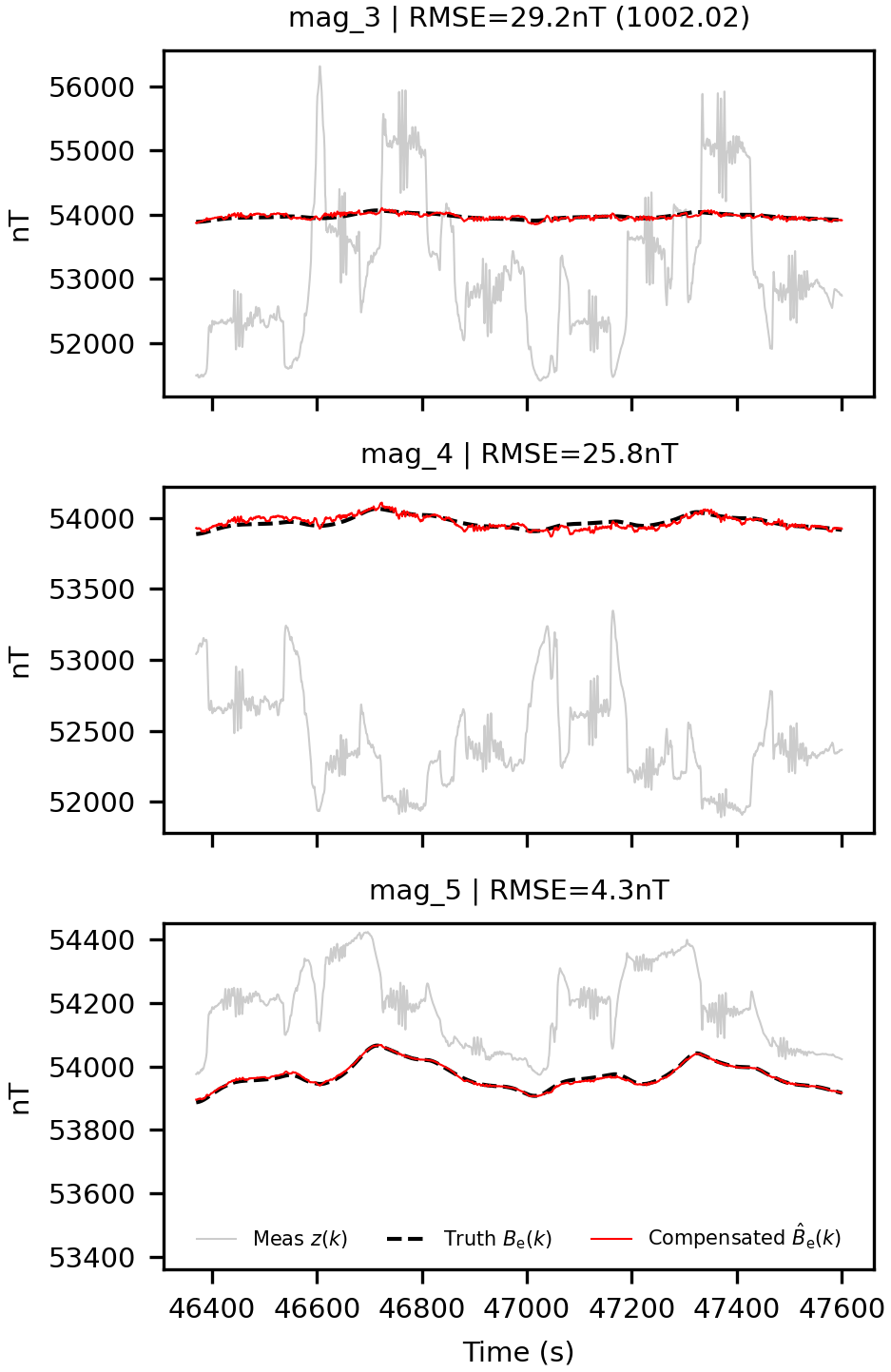}
	\caption{Compensation results for trajectory \texttt{1002.02} using $\hat{\bd{\uptheta}}\bimp$ estimated from trajectory \texttt{1002.02}. The true signal (dashed black line) is identical across all subplots but appears differently due to the different scaling used in each subplot.}
	\label{f4}
\end{figure}

\begin{table}[htbp]
\caption{RMSE Performance Using $\hat{\bd{\uptheta}}\bimp$ Estimated From trajectory \texttt{1002.02}}
\label{tb_calRMSE02}
\centering
\begin{tabular}{lccc}
\hline\hline
\textbf{Flight} & \textbf{mag3\_RMSE} & \textbf{mag4\_RMSE} & \textbf{mag5\_RMSE} \\ 
 \textbf{trajectory}& \textbf{(nT)} & \textbf{(nT)} & \textbf{(nT)} \\ \hline
1002.02 & 29.2  & 25.8  & 4.3  \\
1002.03 & 62.3  & 70.7  & 10.8 \\
1002.04 & 94.8  & 42.8  & 7.6  \\
1002.05 & 93.2  & 54.5  & 8.1  \\
1002.06 & 77.9  & 41.5  & 5.2  \\
1002.07 & 60.2  & 80.0  & 7.1  \\
1002.08 & 52.2  & 100.0 & 9.0  \\
1002.09 & 64.6  & 75.7  & 11.9 \\
1002.10 & 63.0  & 92.2  & 11.0 \\
1002.11 & 54.9  & 107.2 & 13.6 \\
1002.13 & 45.5  & 99.7  & 10.2 \\
1002.15 & 152.5 & 94.6  & 26.2 \\
1002.18 & 133.3 & 137.2 & 22.1 \\
1002.19 & 179.0 & 59.3  & 16.9 \\
1002.20 & 132.2 & 61.8  & 10.2 \\ \hline
\textbf{Average} & \textbf{86.3} & \textbf{76.2} & \textbf{11.6} \\ \hline
\end{tabular}
\end{table}

To visually evaluate the accuracy of the estimated parameter vector $\hat{\bd{\uptheta}}\bimp$, the detailed time-series compensation profiles for the calibration trajectory \texttt{1002.02} are shown in Fig.~\ref{f4}. The figure plots the raw sensor measurements $z(k)$ (gray lines) alongside the reference ground truth $B\be(k)$ (dashed black lines) and the compensated profiles $\hat{B}\be(k)\bimp$ (red lines) for sensors \texttt{mag\_3}, \texttt{mag\_4}, and \texttt{mag\_5}. As shown across all three panels, the raw measurements suffer from extreme aircraft interference, departing from the true external field baseline by several thousand nanoTeslas. However, after applying the estimated parameter vectors via~(\ref{eq_Bimp}), the compensated profiles for all three sensors match the reference ground truth. This high-fidelity alignment demonstrates that the estimation approach successfully identifies the parameter vectors corresponding to each sensor in the figure. 

\begin{figure}[ht]
	\centering
	\includegraphics[width=3.3in]{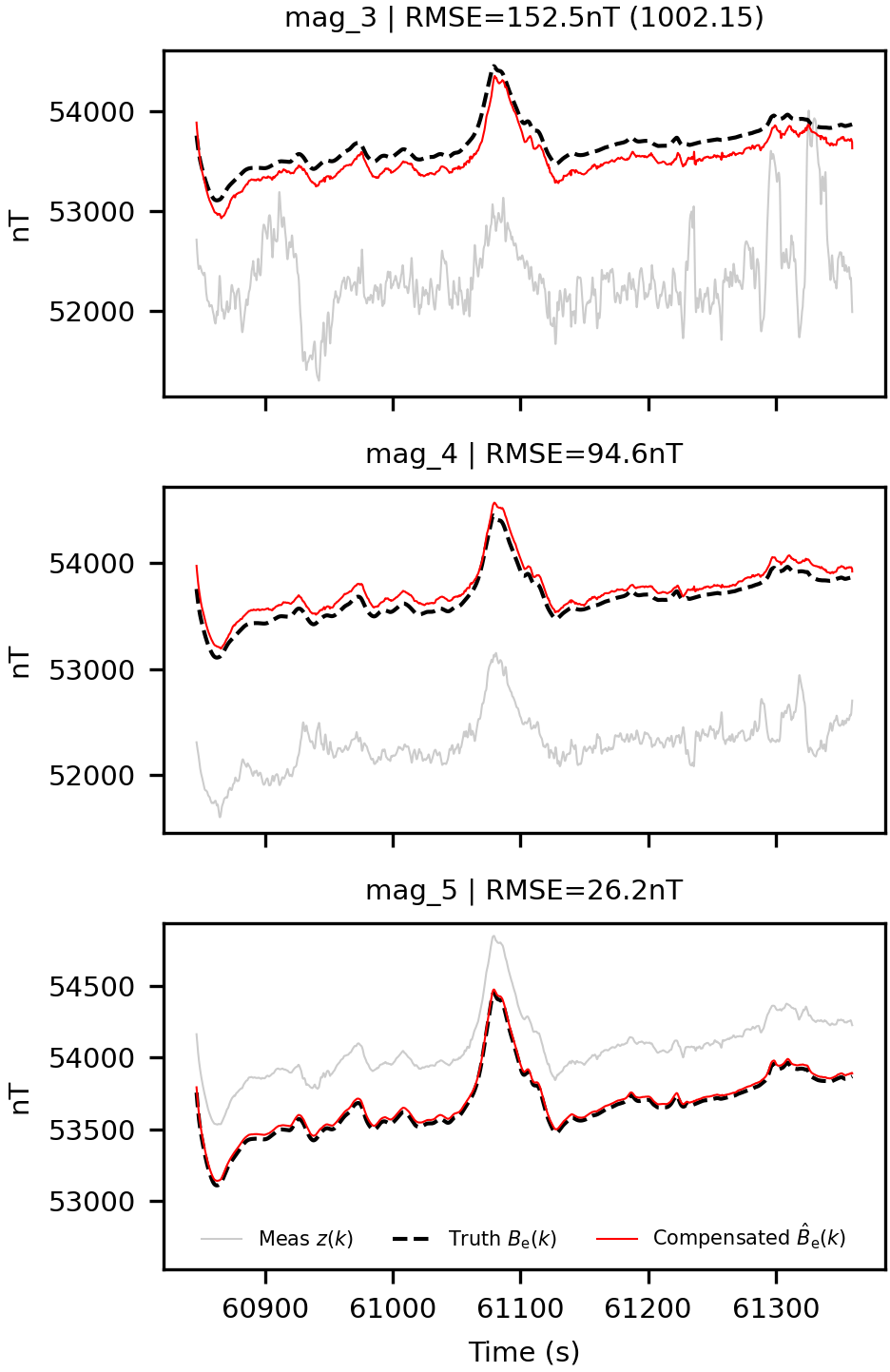}
	\caption{Compensation results for trajectory \texttt{1002.15} using $\hat{\bd{\uptheta}}\bimp$ estimated from trajectory \texttt{1002.02}.}
	\label{f5}
\end{figure}

The compensated RMSE values for all 15 independent evaluation trajectories are summarized in Table~\ref{tb_calRMSE02}, where every trajectory is evaluated using the fixed parameter estimate $\hat{\bd{\uptheta}}\bimp$ obtained from trajectory \texttt{1002.02}. We observe that trajectory \texttt{1002.02} yields the smallest RMSE, while the remaining test trajectories exhibit larger errors. To investigate these performance bounds, the compensation details for trajectory \texttt{1002.15}—which exhibits a significant degradation—are shown in Fig.~\ref{f5}. 
As observed in Fig.~\ref{f5}, the compensated residuals for trajectory \texttt{1002.15} primarily appear as small, bounded biases. This degradation is caused by a combination of physical and mathematical factors. Physically, operational changes in the aircraft platform's environment lead to real-world parameter drift over time. Mathematically, due to severe ill-conditioning during estimation, the parameter vectors overfit the calibration trajectory \texttt{1002.02}. Although this physical drift and mathematical bias cause minor performance degradation across different flight paths, the errors remain tightly bounded, validating the generalizability of the estimated parameters. However, these variations highlight the need for real-time parameter adaptation. This problem serves as the core motivation for an online calibration parameter estimation study, which is beyond the scope of this paper.

\begin{table}[htbp]
\caption{RMSE Performance Using $\hat{\bd{\uptheta}}\bimp$ Estimated From trajectory \texttt{1002.20}}
\label{tb_calRMSE20}
\centering
\begin{tabular}{lccc}
\hline\hline
\textbf{Flight} & \textbf{mag3\_RMSE} & \textbf{mag4\_RMSE} & \textbf{mag5\_RMSE} \\ 
\textbf{trajectory} &\textbf{(nT)} & \textbf{(nT)} & \textbf{(nT)} \\ \hline
1002.02 & 115.2 & 51.4  & 11.4 \\
1002.03 & 93.8  & 61.5  & 17.4 \\
1002.04 & 51.8  & 40.0  & 17.2 \\
1002.05 & 51.3  & 47.4  & 12.6 \\
1002.06 & 51.8  & 28.2  & 11.2 \\
1002.07 & 79.0  & 72.9  & 4.3  \\
1002.08 & 68.5  & 49.0  & 5.9  \\
1002.09 & 66.1  & 56.2  & 7.6  \\
1002.10 & 90.0  & 68.1  & 4.6  \\
1002.11 & 138.2 & 105.9 & 9.2  \\
1002.13 & 64.6  & 52.5  & 6.5  \\
1002.15 & 46.5  & 100.7 & 19.1 \\
1002.18 & 82.8  & 167.6 & 14.1 \\
1002.19 & 66.0  & 85.5  & 11.0 \\
1002.20 & 53.0  & 50.6  & 5.0  \\ \hline
\textbf{Average} & \textbf{74.6} & \textbf{69.2} & \textbf{10.5} \\ \hline
\end{tabular}
\end{table}

\begin{table}[htbp]
\caption{Estimated Parameters (Bias $\beta$ and Permanent Coefficients) by Different Calibration trajectories}
\label{tb_calPara}
\centering
\begin{tabular}{lccccc}
\hline\hline
\textbf{Sensor} & \textbf{Calibration traj.} & \textbf{$\beta$} & \textbf{$a_x$} & \textbf{$a_y$} & \textbf{$a_z$} \\ \hline
\multirow{2}{*}{\texttt{mag\_3}} & 1002.02 & 576.4  & 3814.1 & -2858.9 & -1948.4 \\
                                 & 1002.20 & 1906.8 & 4057.9 & -3251.6 & -2585.7 \\ \hline
\multirow{2}{*}{\texttt{mag\_4}} & 1002.02 & 171.7  & -456.6 & 462.6   & -46.4   \\
                                 & 1002.20 & 2438.5 & 457.1  & 938.9   & -845.2  \\ \hline
\multirow{2}{*}{\texttt{mag\_5}} & 1002.02 & 10.3   & 89.0   & 53.2    & -11.9   \\
                                 & 1002.20 & 103.5  & 164.9  & 234.6   & 2.0     \\ \hline
\end{tabular}
\end{table}

A further study is conducted to evaluate parameter estimation consistency. We use another calibration trajectory \texttt{1002.20} to obtain the parameter vector $\hat{\bd{\uptheta}}\bimp$, which is then evaluated across all trajectories. The resulting RMSE values are listed in Table~\ref{tb_calRMSE20}. The average RMSE values for each sensor remain similar to those reported in Table~\ref{tb_calRMSE02}. However, the RMSE for trajectory \texttt{1002.20} decreases, while the error for trajectory \texttt{1002.02} increases significantly. This shift demonstrates that the estimate $\hat{\bd{\uptheta}}\bimp$ overfits the specific calibration dataset, which is typical for ill-conditioned problems. 

To illustrate this variation, the first four estimated parameters of $\hat{\bd{\uptheta}}\bimp$ obtained from trajectories \texttt{1002.02} and \texttt{1002.20} are compared in Table~\ref{tb_calPara}. The parameter estimate values obtained from different flight trajectories differ substantially, confirming that the parameter estimation itself is highly inconsistent due to the ill-conditioned nature of the problem. Fortunately, the forward evaluation model acts as the inverse process of this unstable estimation. Although the individual parameters within $\hat{\bd{\uptheta}}\bimp$ contain significant estimation errors, the forward model computation inherently suppresses these instabilities. This reconstruction process drastically limits the impact of parameter inaccuracies on the final field compensation.

In summary, the parameter estimation framework successfully achieves the calibration parameter estimation, provided at least a vector sensor and a reliable anomaly map are available, and the path remains isolated from dynamic magnetic objects not included in the map (unless a scalar sensor is mounted away from the platform). Furthermore, while severe ill-conditioning yields large estimation errors within $\hat{\bd{\uptheta}}\bimp$, the forward compensation process inherently suppresses these instabilities, preserving overall accuracy. Ultimately, real-time calibration parameter estimation remains necessary to mitigate parameter overfitting, possible real-world drift and sudden change of magnetic environment inside the platform.

\subsection{External Field $B\be$ Dynamic Estimation Test}\label{s6b}

This section evaluates the performance of the KFF algorithm developed in Section~\ref{s5} using the DAF-MIT MagNav dataset. The evaluation utilizes the \texttt{flux\_b} sensor to obtain the unit vector $\bd{u}\ts$ and its numerical derivative $\dot{\bd{u}}\ts$, which serve as inputs to the dynamic model. The multi-sensor suite selected for fusion comprises the \texttt{mag\_3}, \texttt{mag\_4}, and \texttt{mag\_5} magnetometers. The calibration parameters for these sensors are estimated \emph{a priori} using flight trajectory \texttt{1002.02} in Section~\ref{s6a}, and the subsequent performance tests are conducted across all 15 flight trajectory=ies. To assess estimation accuracy, the error between the estimated $\hat B\be(k)\bKFF$ and the ground truth $B\be(k)$ provided by \texttt{mag\_1\_c} is evaluated. 

During the test, the time interval $\Delta t$ is 0.1s, the process noise covariance matrix is set as $\bd Q = \mathrm{diag}([0.1, 0.01])$, and the measurement noise variances for \texttt{mag\_3}, \texttt{mag\_4}, and \texttt{mag\_5} are set as $R_1=900$, $R_2=500$, and $R_3=10$, respectively.
The initial state is set as $\bd x(1) = [z_1(1)\ \ 0.0]'$, and its error covariance $\bd P(1) = \mathrm{diag}([900.0\ \ 0.1])$.

For benchmarking purposes, the classical TL model is also evaluated. Its real-time estimate, $\hat{B}\be(k)\bTL$, is derived from~(\ref{eq_TL_eval}) by removing the non-real-time term $\overline{\bd A\bTL \bd{\uptheta}\bTL}$ as
\begin{equation}
\hat B\be(k)\bTL = z(k) - \bd A\bTL(k)\hat{\bd{\uptheta}}\bTL
\end{equation}
In addition, the evaluation approach with the augmented parameter introduced in Section~\ref{s6a} is included in the comparison and designated as AUG, with its corresponding estimate $\hat B\be(k)\bimp$ given in~(\ref{eq_Bimp}). It should be noted that the AUG approach is inherently impractical for dynamic $B\be(k)$ estimation. This limitation arises because the reference magnetometer \texttt{mag\_1} cannot be deployed in real operational scenarios, which leads to an absence of an accurate $B\be(k)$ when forming $\bd A\bimp(k)$. Although static anomaly maps, the IGRF, and diurnal variations can approximate the external field when dynamic interferences are absent, this approximation fails to account for dynamic objects. The AUG approach is therefore included strictly for comparative analysis. Since it uses the ground truth $B\be(k)$ to form $\bd A\bimp(k)$, it provides nearly optimal estimates for a single sensor when the errors of the parameter vector and $\bd u\ts$ are negligible.

\begin{table}[htbp]
\caption{Performance Comparison of dynamic $B\be$ estimation approaches}
\label{tb_dyn_tol}
\centering
\begin{tabular}{lccc}
\hline\hline
\textbf{Approach} & Bias & \textbf{Err. STDV} & \textbf{RMSE} \\ 
 & \textbf{(nT)} & \textbf{(nT)} & \textbf{(nT)} \\ \hline
TL            & 356.9 & 28.2 & 359.2 \\
AUG & 42.7  & 31.8 & 58.0  \\
\textbf{KFF}    & \textbf{10.0}  & \textbf{5.9}  & \textbf{12.3}  \\ \hline
\end{tabular}
\end{table}

Table~\ref{tb_dyn_tol} presents the performance summary of the dynamic $B\be$ estimation approaches across the evaluation flight trajectoties. The detailed results of them are presented in Tables~\ref{tb_tl}--\ref{tb_kff3} in Appendix~\ref{a1}, respectively. The evaluation metrics comprise the bias error, the error standard deviation (after removing the bias), and the total RMSE. Unlike KFF, which provides a fused estimate from all three sensors, the TL and AUG approaches can only generate estimates on a sensor-by-sensor basis; hence, their reported results are averaged across the individual sensors. The findings demonstrate that the proposed KFF significantly outperforms both the classical TL model and the idealized AUG benchmark across all criteria. Specifically, the classical TL approach exhibits a severe bias error of 356.9~nT, leading to a large overall RMSE of 359.2~nT. Meanwhile, the optimal single-sensor AUG estimator demonstrates inferior performance compared to KFF, primarily because it lacks the benefits of multi-sensor fusion.

\begin{table}[htbp]
\caption{Performance Comparison under Different Sensor Configurations}
\label{tb_KFF_comp}
\centering
\begin{tabular}{llccc}
\hline\hline
\textbf{Approach} & \textbf{Sensor} & \textbf{Bias} & \textbf{Err. STDV} & \textbf{RMSE} \\ 
 & & \textbf{(nT)} & \textbf{(nT)} & \textbf{(nT)} \\ \hline
AUG & \texttt{mag\_3} & 57.9 & 51.2 & 86.3 \\
 & \texttt{mag\_4} & 60.7 & 38.9 & 76.2 \\
 & \texttt{mag\_5} & 9.5 & 5.3 & 11.6 \\ \hline
KFF & \texttt{mag\_3} & 60.8 & 54.2 & 90.6 \\
 & \texttt{mag\_4} & 62.4 & 45.2 & 80.9 \\
 & \texttt{mag\_5} & 9.5 & 5.4 & 11.6 \\
 & \texttt{mag\_3,4} & 38.1 & 43.1 & 60.5 \\
 & \texttt{mag\_3,4,5} & 10.0 & 5.9 & 12.3 \\ \hline\hline
\end{tabular}
\end{table}

Table \ref{tb_KFF_comp} provides a performance comparison under different sensor 
configurations to evaluate the impact and structural robustness of the proposed KFF. 
Several key insights can be extracted from these results. First, when utilizing a 
single sensor, the KFF tracking accuracy remains remarkably close to the idealized 
AUG benchmark. The AUG approach uses the true external field baseline from 
\texttt{mag\_1\_c} to construct its matrix $\bd A\bimp(k)$, and the KFF achieves 
comparable accuracy without this information. 

More importantly, the three-sensor KFF configuration ($\text{mag\_3, 4, 5}$) demonstrates 
robustness. Even when the clean $\text{mag\_5}$ sensor (individual RMSE = 11.6~nT) 
is simultaneously fused alongside $\text{mag\_3}$ and $\text{mag\_4}$—both of which 
suffer from heavy platform dynamic disturbances (RMSE values of 90.6~nT and 80.9~nT, 
respectively)—the KFF successfully prevents these high-error channels from corrupting 
the estimation. The resulting fused error stabilizes at 12.3~nT, providing asymptotic 
 performance bound of the single best sensor. This minimal variance of only 
0.7~nT under highly asymmetric noise exposure demonstrates that the filter effectively 
isolates severe, localized aircraft dynamic anomalies. 

Furthermore, when fusing multiple sensors characterized by equivalent error baselines, 
such as the dual-sensor $\text{mag\_3, 4}$ configuration, the KFF framework exhibits 
a high variance-reduction efficacy. The dual-sensor fused result yields a combined 
RMSE of 60.5~nT, providing a significant accuracy optimization over both individual 
sensors. Collectively, these empirical findings demonstrate that the proposed KFF 
tracks the state and fuses multi-sensor measurements robustly.

\begin{figure}[ht]
	\centering
	\includegraphics[width=3.3in]{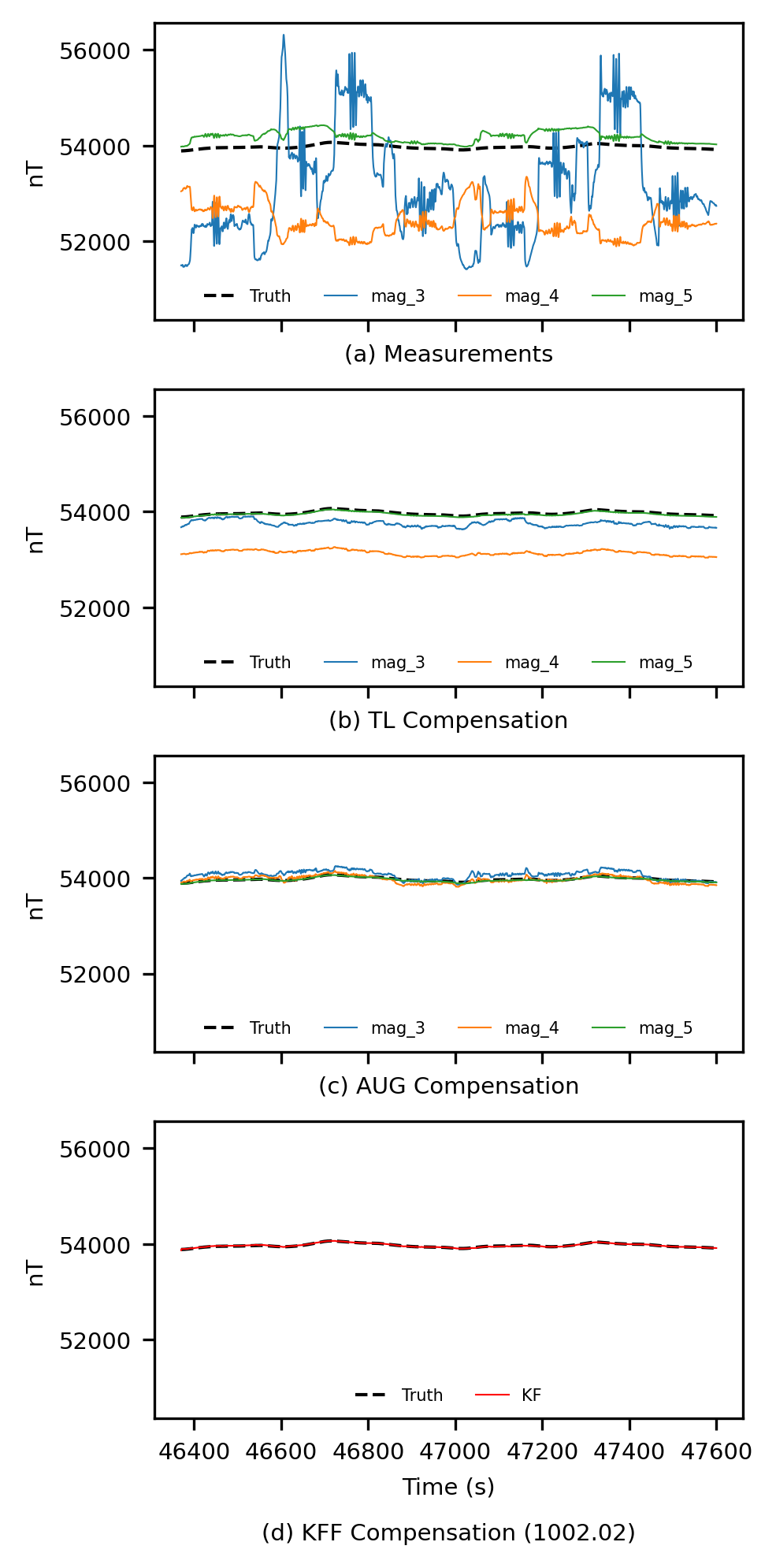}
	\caption{Dynamic estimation results under flight trajectory \texttt{1002.02} (calibration traj.).} 
	\label{f6}
\end{figure}

\begin{figure}[ht]
	\centering
	\includegraphics[width=3.5in]{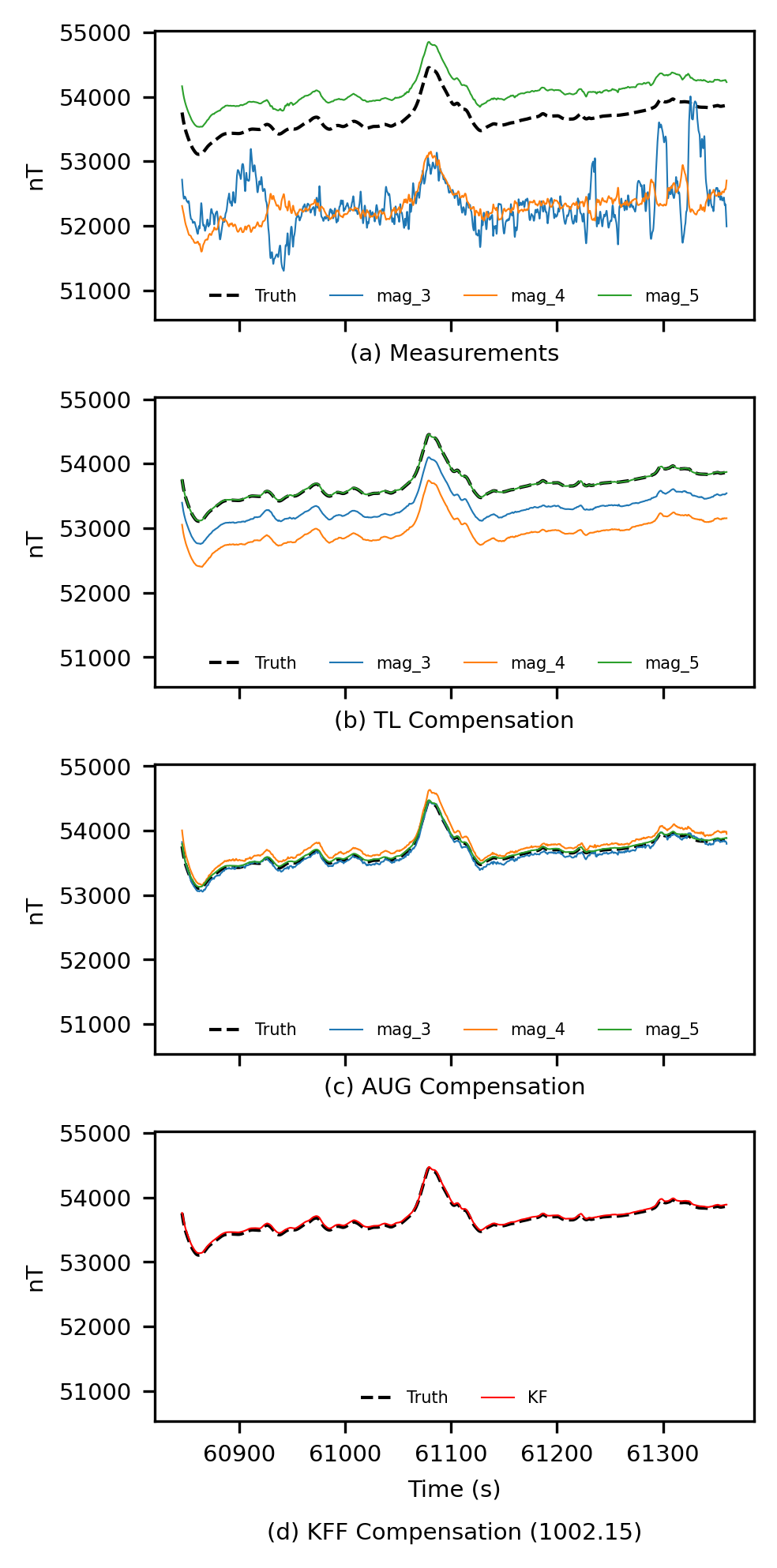}
	\caption{Dynamic estimation results under flight trajectory \texttt{1002.15}}
	\label{f7}
\end{figure}

To visually evaluate the tracking performance, the time-series comparisons of the dynamic $B\be$ estimation approaches for flight trajectories \texttt{1002.02} and \texttt{1002.15} are shown in Figs.~\ref{f6} and~\ref{f7}, respectively, where \texttt{1002.02} serves as the calibration trajectory and \texttt{1002.15} represents the case with the largest RMSE. Subfigure~(a) displays the raw, uncompensated measurements, highlighting severe aircraft-induced dynamic interferences and substantial offsets across \texttt{mag\_3}, \texttt{mag\_4}, and \texttt{mag\_5} relative to the ground truth. As shown in subfigure~(b), the classical TL model removes high-frequency variations but fails to eliminate large, persistent bias errors under real-time constraints, leaving the sensor tracks vertically separated from the truth line. The idealized AUG compensation in subfigure~(c) significantly closes this gap, yet minor tracking deviations remain due to its single-sensor limitation, which is particularly visible in Fig.~\ref{f7}(c). In contrast, the proposed KFF in subfigure~(d) achieves the highest estimation accuracy; by executing effective multi-sensor information fusion, the KFF estimate tracks the ground truth almost perfectly, effectively eliminating both the persistent biases and dynamic fluctuations across both flight trajectories.

\section{Other Possible Approaches}\label{s7}

In addition to the linearized approach proposed in Section~\ref{s4}, several alternative methods were developed during the course of this research. While these approaches are theoretically attractive, experimental results show that none outperform the proposed linearized approach. This section summarizes these methods and discusses the reasons for their inferior performance.

\subsection{Nonlinear Model Using the Iterated Least Squares Estimator}

As discussed previously, the ILS algorithm can be applied directly to the nonlinear model in~(\ref{eq_modelScalar}) to estimate $\bd{\uptheta}\bimp$. The primary advantage of this approach is that it avoids the approximation error introduced by linearization. However, for an ill-conditioned estimation problem, the repeated matrix inversion required by ILS can increase numerical instability and make the solution highly sensitive to measurement perturbations.

\begin{figure}[ht]
	\centering
	\includegraphics[width=3.5in]{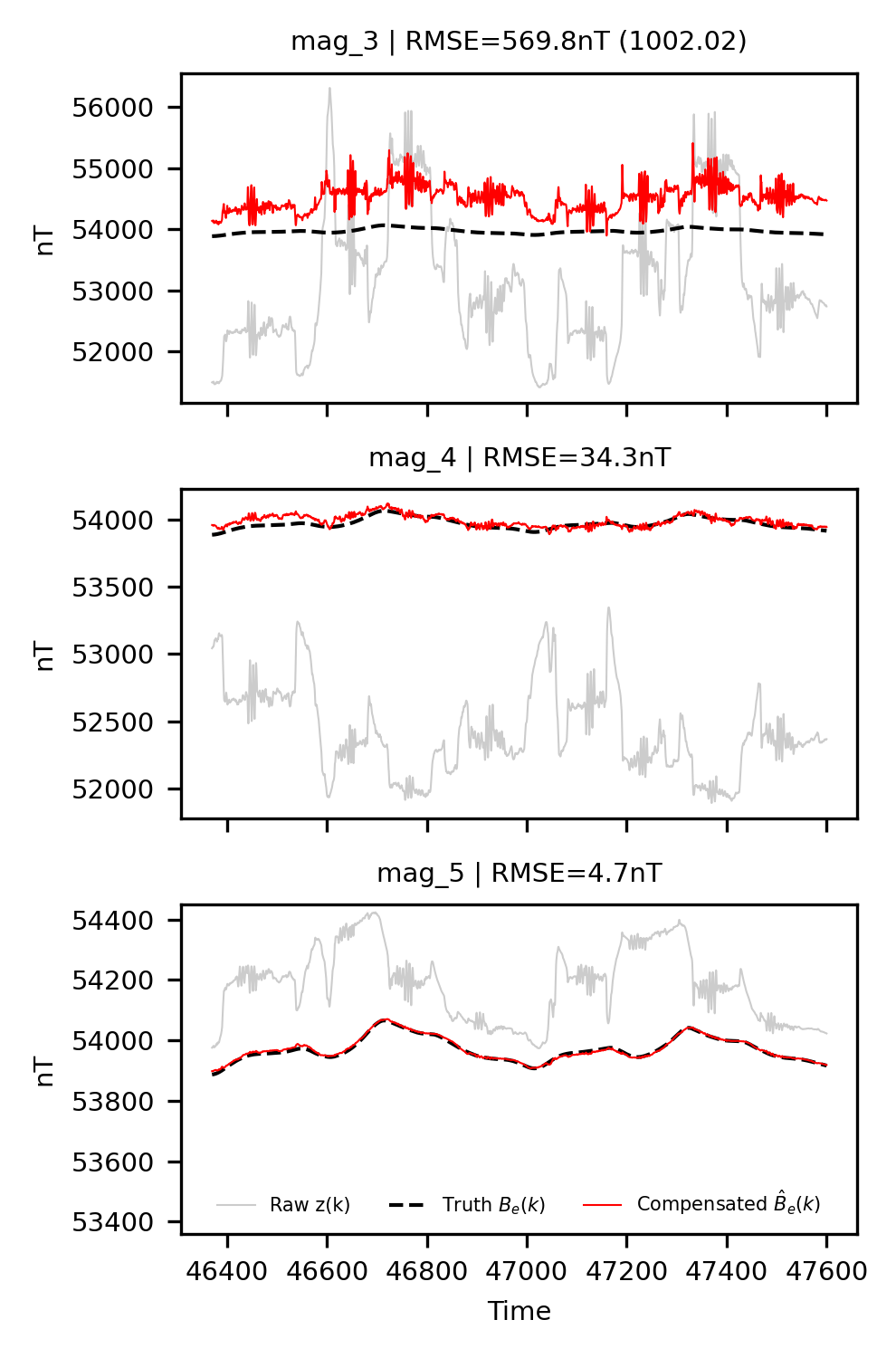}
	\caption{Iterated least squares estimation results for flight trajectory \texttt{1002.02}.}
	\label{f8}
\end{figure}

\begin{table}[htbp]
\centering
\caption{Measurement residual RMS and compensated RMSE comparison.}
\label{tb_ils}
\begin{tabular}{lcccc}
\hline
\textbf{Sensor} &
\textbf{1st iter. res.} &
\textbf{3rd iter. res.} &
\textbf{LS RMSE} &
\textbf{ILS RMSE} \\
&
\textbf{RMS (nT)}&
\textbf{RMS (nT)}&
\textbf{(nT)}&
\textbf{(nT)}\\
\hline
\texttt{mag\_3} & 1655 & 88 & 29.2 & 569.8 \\
\texttt{mag\_4} & 143 & 27 & 25.8 & 34.3 \\
\texttt{mag\_5} & 7 & 4 & 4.3 & 4.7 \\
\hline
\end{tabular}
\end{table}

The nonlinear model is expressed as

\begin{equation}
    \bd Z = \bd H_{\mathrm{ILS}}(\bd{\uptheta}\bimp) + \bd V
\end{equation}
This is the stacked form of~(\ref{eq_modelScalar}) from time 1 to $n$. The parameter vector $\bd{\uptheta}\bimp$ is estimated iteratively~\cite{BarShalom2001} by

\begin{equation}
    \hat{\bd{\uptheta}}\bimp^{j+1} = \hat{\bd{\uptheta}}\bimp^{j} +
    (\bd J^{T}\bd J)^{-1} \bd J^{T} \left(\bd Z - \hat{\bd Z}^{j}
    \right)
\end{equation}
where $j$ denotes the iteration index. The Jacobian matrix $\bd J$ above is given by

\begin{equation}
    \bd J=\left.\frac{\partial \bd H_{\mathrm{ILS}}(\cdot)}
         {\partial \bd{\uptheta}\bimp}
    \right|_{\bd{\uptheta}\bimp = \hat{\bd{\uptheta}}\bimp^{j}}
\end{equation}
and
\begin{equation}
    \hat{\bd Z}^{j}=\bd H_{\mathrm{ILS}}(\hat{\bd{\uptheta}}\bimp^{j})
\end{equation}

The ILS compensation results, using the same flight trajectory, \texttt{1002.02}, are shown in Fig.~\ref{f8}, which was obtained under the same conditions as the LS results in Fig.~\ref{f4}. The measurement residual RMS obtained during the ILS iterations and the compensated RMSE results for both LS and ILS are summarized in Table~\ref{tb_ils}. It can be seen that the ILS solution performs noticeably worse than the LS solution. Although ILS successfully reduces the measurement residual RMS, the improved measurement fitting does not translate into improved compensation accuracy. This apparent contradiction is a consequence of the severe ill-conditioning of the estimation problem. The condition number of the Jacobian matrix is on the order of $10^{7}$, indicating that the parameter vector $\bd{\uptheta}\bimp$ is poorly observable. In such a situation, substantially different parameter vectors can produce similar measurement predictions. Consequently, the repeated inversion of $(\bd J^{T}\bd J)^{-1}$ amplifies the sensitivity of the solution to measurement perturbations and numerical errors. As shown in Table~\ref{tb_ils}, a significant reduction in the measurement residual RMS is accompanied by degraded compensation accuracy. These results indicate that the nonlinear iterations increasingly fit the particular measurement dataset used for estimation rather than improving the accuracy of the estimated parameters. Therefore, despite eliminating the linearization approximation, the nonlinear ILS estimator yields inferior compensation performance for this highly ill-conditioned problem. The linear LS solution therefore provides a more robust and reliable estimate.

It should be noted that the ill-conditioning of the estimation problem is mainly determined by the maneuvering profile of the flight trajectory rather than by the choice of estimation algorithm. Consequently, both the LS and ILS solutions are affected by the poor observability of the calibration parameters, although the repeated matrix inversions in ILS make it more sensitive to this issue. Despite the extensive maneuvers performed in flight trajectory \texttt{1002.02}, which are consistent with conventional calibration guidelines, the resulting regression matrix still exhibits a condition number on the order of $10^{7}$. This indicates that the calibration parameters remain poorly observable under the current excitation conditions. Future work should therefore focus on improving parameter observability through optimized flight-path design, such as incorporating larger altitude changes or other maneuvers that excite the calibration parameters more effectively.

\subsection{Vector Model for Calibration Parameter Estimation}

When a vector magnetometer is available, its measurement vector can be used directly to construct a linear model. Unlike the scalar formulation proposed in Section~\ref{s4}, the vector model incorporates all three magnetic-field components simultaneously. As a result, the number of available observations is increased, potentially providing additional information for parameter estimation. In addition, the vector formulation does not require the linearization used in the scalar model and is therefore free from linearization error. This section develops the corresponding vector model and evaluates its estimation performance.

\begin{figure}[ht] 
\centering 
\includegraphics[width=3.5in]{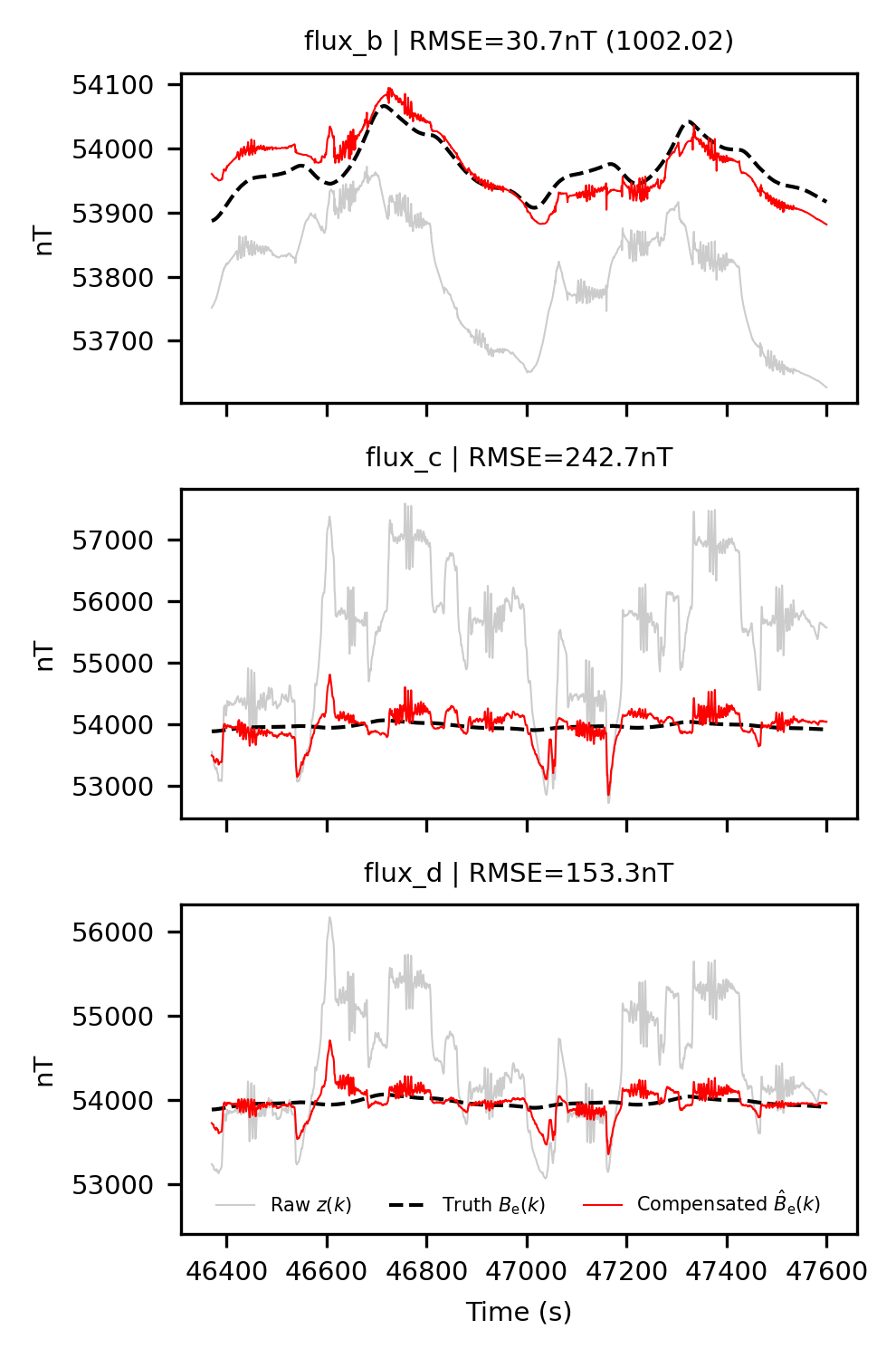} 
\caption{Vector model LS results for the flight trajectory \texttt{1002.02}.} \label{f9} 
\end{figure} 

\begin{figure}[ht] 
\centering 
\includegraphics[width=3.5in]{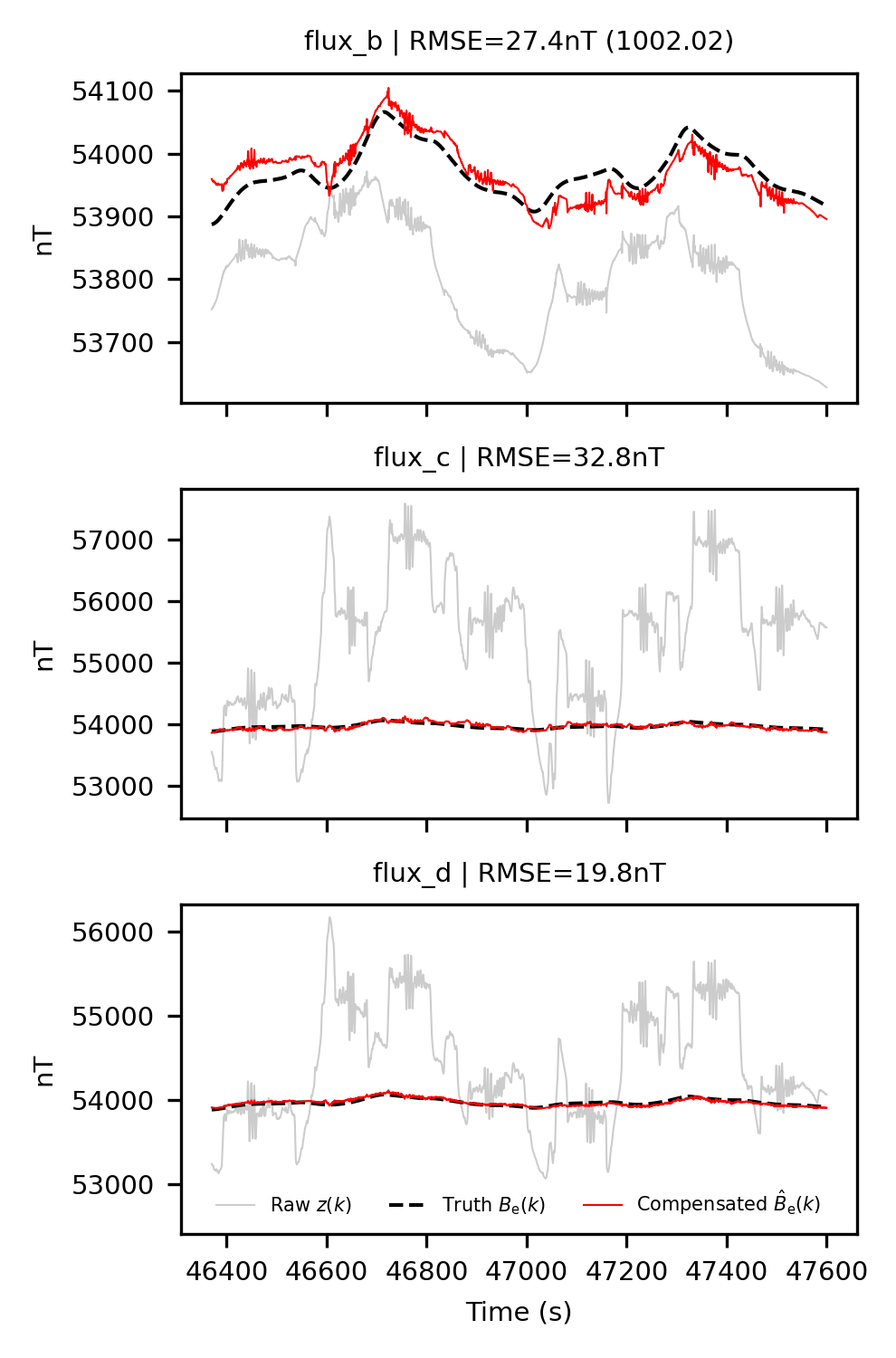} 
\caption{Scalar model on the vector norm LS results for the flight trajectory \texttt{1002.02}.} \label{f10} 
\end{figure} 

The measurement vector at time $k$ is 
\begin{equation} 
    \bd z(k) = [B\txs(k)\ \ B\tys(k)\ \ B\tzs(k)]'
\end{equation} 
The linear model is given in~(\ref{eq_modelV}). Stacking the measurements from time $1$ to $n$ yields
\begin{equation} 
\bd Z_{\mathrm{v}} = \bd H_{\mathrm{v}}\,\bd{\uptheta}\bimp + \bd V_{\mathrm{v}}
\end{equation} 
The parameter vector is estimated using regularized LS, \begin{equation}\label{eq_LS3} 
\hat{\bd{\uptheta}}\bimp = \left(\bd H_{\mathrm{v}}' \bd H_{\mathrm{v}} + \lambda \mathbf{I}\right)^{-1} \bd H_{\mathrm{v}}' \bd Z_{\mathrm{v}}
\end{equation} 

\begin{table}[ht]
\centering
\caption{RMSE comparison between the vector model and the scalar model (proposed in Section~\ref{s4}) for flight trajectory \texttt{1002.02}.}
\label{tb_vmodel}
\begin{tabular}{lccc}
\hline
\textbf{Sensor} & \textbf{Vector model} & \textbf{Scalar model} & \textbf{Error reduction} \\
       & \textbf{RMSE (nT)}    & \textbf{RMSE (nT)}    & \textbf{(\%)} \\
\hline
\texttt{flux\_b} & 30.7  & 27.4 & 10.7 \\
\texttt{flux\_c} & 242.7 & 32.8 & 86.5 \\
\texttt{flux\_d} & 153.3 & 19.8 & 87.1 \\
\hline
\end{tabular}
\end{table}

The vector-model LS estimator was tested on flight trajectory \texttt{1002.02} using the three fluxgate sensors \texttt{flux\_b}, \texttt{flux\_c}, and \texttt{flux\_d}. The results are shown in Fig.~\ref{f9}. For comparison, the approach proposed in Section~\ref{s4}, which applies the scalar linearized model to the norm of $\bd z(k)$, was also evaluated. The corresponding results are shown in Fig.~\ref{f10}. Table~\ref{tb_vmodel} summarizes the RMSEs obtained by the two methods. The scalar model consistently outperforms the vector model. The performance gain is particularly significant for \texttt{flux\_c} and \texttt{flux\_d}, where the RMSE is reduced by more than 86\%.

Although the vector model eliminates linearization error, the scalar model achieves markedly superior compensation performance on this dataset. This discrepancy is largely driven by numerical stability: the matrix condition number for the vector model rises $10^8$, compared to $10^7$ for the scalar model. In addition, the individual vector channels of $\mathbf{z}(k)$ exhibit higher noise levels than its total field norm, which further decreases the vector model's estimation accuracy. Consequently, despite the presence of residual linearization error, the linearized scalar model yields substantially lower compensated RMSEs across all three sensors.




\section{Conclusions}\label{s8}
In this paper, a comprehensive, two-stage aeromagnetic compensation and dynamic tracking framework has been developed that overcomes the structural and real-time limitations of the classical TL methodology. In Stage one, an augmented linear model for sensor-based calibration parameter estimation is established, overcoming traditional TL limitations including the information distortion caused by the band-pass filter and the omission of sensor measurement biases. In Stage 2, leveraging these pre-estimated parameters, a map-less KFF approach is developed to dynamically estimate the external magnetic field from corrupted multi-sensor observations. This achieves real-time joint dynamic estimation and multi-sensor compensation without relying on prior geomagnetic maps or external reference fields.

Experimental validations on the DAF-MIT MagNav flight dataset led to several key insights:
\begin{itemize}
    \item \textbf{Suppression of Parameter Instabilities:} Although the parameter estimation problem remains highly ill-conditioned (with condition numbers on the order of $10^7$ due to collinearity and trajectory limitations), the forward compensation model acts as an inverse process that naturally suppresses parameter instabilities, preserving high overall compensation accuracy.
    \item \textbf{Robust Multi-Sensor Information Fusion:} The KFF algorithm demonstrates remarkable structural robustness. When fusing clean sensors alongside highly corrupted, high-noise channels, the filter effectively isolates localized dynamic anomalies, ensuring the fused estimate asymptotically approaches the performance bound of the cleanest sensor. When fusing sensors with equivalent noise baselines, the framework yields substantial variance reduction and accuracy optimization over individual sensors.
    \item \textbf{Superiority of Linearized Scalar Calibration:} Linearized scalar calibration consistently outperforms both the non-linear scalar model and the three-axis vector model. The non-linear model relies on iterative matrix inversions over an ill-conditioned regressor, yielding unstable parameter estimates, while the vector model exhibits even poorer observability and higher measurement noise.
\end{itemize}

Future work will focus on extending this framework to incorporate real-time online calibration parameter estimation. This is essential to address parameter overfitting to the calibration data as well as parameter drifting over long flight durations. Additionally, optimized flight-path trajectories designed to excite the calibration parameters more effectively will be investigated, thereby directly improving parameter observability and reducing the underlying ill-conditioning of the estimation problem.

\appendices
\section{Detailed Results of the KFF}\label{a1}

\begin{table}[htbp]
\caption{Dynamic $B\be$ Estimation Performance of TL Approach}
\label{tb_tl}
\centering
\begin{tabular}{llrrr}
\hline\hline
\textbf{Sensor} & \textbf{Flight} & \textbf{Bias} & \textbf{Err. STDV} & \textbf{RMSE} \\ 
 & \textbf{trajectory}& \textbf{(nT)} & \textbf{(nT)} & \textbf{(nT)} \\ \hline
\texttt{mag\_3} & 1002.02 & -213.6 & 63.1 & 222.8 \\
 & 1002.03 & -248.6 & 75.7 & 259.8 \\
 & 1002.04 & -271.0 & 56.7 & 276.9 \\
 & 1002.05 & -294.2 & 34.8 & 296.3 \\
 & 1002.06 & -282.2 & 38.0 & 284.8 \\
 & 1002.07 & -243.7 & 38.9 & 246.8 \\
 & 1002.08 & -236.3 & 32.7 & 238.6 \\
 & 1002.09 & -222.4 & 24.8 & 223.8 \\
 & 1002.10 & -204.1 & 46.6 & 209.4 \\
 & 1002.11 & -190.9 & 46.2 & 196.4 \\
 & 1002.13 & -213.5 & 24.8 & 215.0 \\
 & 1002.15 & -354.6 & 17.1 & 355.0 \\
 & 1002.18 & -374.3 & 71.9 & 381.1 \\
 & 1002.19 & -410.2 & 26.7 & 411.1 \\
 & 1002.20 & -328.6 & 88.7 & 340.4 \\ \cline{2-5}
 & \textbf{Sub Avg} & \textbf{272.5} & \textbf{45.8} & \textbf{277.2} \\ \hline
\texttt{mag\_4} & 1002.02 & -832.1 & 38.8 & 833.0 \\
 & 1002.03 & -874.2 & 58.0 & 876.1 \\
 & 1002.04 & -849.4 & 18.5 & 849.6 \\
 & 1002.05 & -863.9 & 25.6 & 864.3 \\
 & 1002.06 & -802.0 & 16.7 & 802.2 \\
 & 1002.07 & -767.4 & 18.6 & 767.6 \\
 & 1002.08 & -763.0 & 30.5 & 763.6 \\
 & 1002.09 & -750.3 & 21.0 & 750.6 \\
 & 1002.10 & -743.4 & 19.6 & 743.7 \\
 & 1002.11 & -712.0 & 41.8 & 713.3 \\
 & 1002.13 & -738.7 & 22.2 & 739.1 \\
 & 1002.15 & -719.4 & 20.6 & 719.7 \\
 & 1002.18 & -706.3 & 67.3 & 709.4 \\
 & 1002.19 & -782.6 & 34.9 & 783.4 \\
 & 1002.20 & -833.9 & 59.7 & 836.0 \\ \cline{2-5}
 & \textbf{Sub Avg} & \textbf{782.6} & \textbf{32.9} & \textbf{783.4} \\ \hline
\texttt{mag\_5} & 1002.02 & -22.3  & 5.7  & 23.0  \\
 & 1002.03 & -28.6  & 8.0  & 29.7  \\
 & 1002.04 & -26.0  & 5.7  & 26.7  \\
 & 1002.05 & -23.9  & 5.7  & 24.6  \\
 & 1002.06 & -20.5  & 5.5  & 21.2  \\
 & 1002.07 & -16.6  & 4.4  & 17.2  \\
 & 1002.08 & -17.2  & 2.4  & 17.4  \\
 & 1002.09 & -13.3  & 8.5  & 15.8  \\
 & 1002.10 & -13.1  & 6.4  & 14.6  \\
 & 1002.11 & -11.2  & 8.3  & 14.0  \\
 & 1002.13 & -14.3  & 4.3  & 14.9  \\
 & 1002.15 & 3.8    & 3.8  & 5.4   \\
 & 1002.18 & -1.2   & 4.4  & 4.5   \\
 & 1002.19 & -6.5   & 8.4  & 10.6  \\
 & 1002.20 & -13.9  & 6.7  & 15.5  \\ \cline{2-5}
 & \textbf{Sub Avg} & \textbf{15.5}  & \textbf{5.9}  & \textbf{17.0}  \\ \hline
\textbf{Total Avg} & & \textbf{356.9} & \textbf{28.2} & \textbf{359.2} \\ \hline\hline
\end{tabular}
\end{table}

\begin{table}[htbp]
\caption{Dynamic $B\be$ Estimation Performance of AUG Approach}
\label{tb_imp}
\centering
\begin{tabular}{llrrr}
\hline\hline
\textbf{Sensor} & \textbf{Flight} & \textbf{Bias} & \textbf{Err. STDV} & \textbf{RMSE} \\ 
 & \textbf{trajectory}& \textbf{(nT)} & \textbf{(nT)} & \textbf{(nT)} \\ \hline
\texttt{mag\_3} & 1002.02 & 0.0 & 29.2 & 29.2 \\
 & 1002.03 & -30.1 & 54.5 & 62.3 \\
 & 1002.04 & -89.9 & 29.9 & 94.8 \\
 & 1002.05 & -80.7 & 46.5 & 93.2 \\
 & 1002.06 & -52.1 & 57.9 & 77.9 \\
 & 1002.07 & -20.1 & 56.7 & 60.2 \\
 & 1002.08 & 11.1 & 51.1 & 52.2 \\
 & 1002.09 & -25.5 & 59.3 & 64.6 \\
 & 1002.10 & -0.4 & 63.0 & 63.0 \\
 & 1002.11 & 4.0 & 54.7 & 54.9 \\
 & 1002.13 & -6.8 & 45.0 & 45.5 \\
 & 1002.15 & -146.7 & 41.8 & 152.5 \\
 & 1002.18 & -116.7 & 64.5 & 133.3 \\
 & 1002.19 & -174.2 & 40.9 & 179.0 \\
 & 1002.20 & -110.6 & 72.4 & 132.2 \\ \cline{2-5}
 & \textbf{Sub Avg} & \textbf{57.9} & \textbf{51.2} & \textbf{86.3} \\ \hline
\texttt{mag\_4} & 1002.02 & 0.0 & 25.8 & 25.8 \\
 & 1002.03 & -39.2 & 58.9 & 70.7 \\
 & 1002.04 & -37.5 & 20.6 & 42.8 \\
 & 1002.05 & -33.6 & 42.9 & 54.5 \\
 & 1002.06 & 28.3 & 30.3 & 41.5 \\
 & 1002.07 & 70.4 & 38.0 & 80.0 \\
 & 1002.08 & 94.4 & 33.1 & 100.0 \\
 & 1002.09 & 64.2 & 40.2 & 75.7 \\
 & 1002.10 & 80.0 & 45.8 & 92.2 \\
 & 1002.11 & 94.3 & 51.0 & 107.2 \\
 & 1002.13 & 97.8 & 19.6 & 99.7 \\
 & 1002.15 & 91.0 & 25.8 & 94.6 \\
 & 1002.18 & 121.8 & 63.3 & 137.2 \\
 & 1002.19 & 53.1 & 26.3 & 59.3 \\
 & 1002.20 & -4.7 & 61.7 & 61.8 \\ \cline{2-5}
 & \textbf{Sub Avg} & \textbf{60.7} & \textbf{38.9} & \textbf{76.2} \\ \hline
\texttt{mag\_5} & 1002.02 & 0.0 & 4.3 & 4.3 \\
 & 1002.03 & -6.5 & 8.7 & 10.8 \\
 & 1002.04 & -5.1 & 5.7 & 7.6 \\
 & 1002.05 & -1.7 & 7.9 & 8.1 \\
 & 1002.06 & 3.3 & 4.1 & 5.2 \\
 & 1002.07 & 6.5 & 3.0 & 7.1 \\
 & 1002.08 & 8.8 & 2.3 & 9.0 \\
 & 1002.09 & 9.2 & 7.6 & 11.9 \\
 & 1002.10 & 10.2 & 4.1 & 11.0 \\
 & 1002.11 & 10.4 & 8.9 & 13.6 \\
 & 1002.13 & 9.9 & 2.4 & 10.2 \\
 & 1002.15 & 25.9 & 3.8 & 26.2 \\
 & 1002.18 & 21.7 & 4.3 & 22.1 \\
 & 1002.19 & 15.6 & 6.5 & 16.9 \\
 & 1002.20 & 8.0 & 6.3 & 10.2 \\ \cline{2-5}
 & \textbf{Sub Avg} & \textbf{9.5} & \textbf{5.3} & \textbf{11.6} \\ \hline
\textbf{Total Avg} & & \textbf{42.7} & \textbf{31.8} & \textbf{58.0} \\ \hline\hline
\end{tabular}
\end{table}

\begin{table}[htbp]
\caption{Dynamic $B\be$ Estimation Performance of the KFF Approach (Fusion of \texttt{mag\_3, mag\_4 and mag\_5})}
\label{tb_kff3}
\centering
\begin{tabular}{lrrr}
\hline\hline
\textbf{Flight} & \textbf{Bias} & \textbf{Err. STDV} & \textbf{RMSE} \\
\textbf{trajectory} & \textbf{(nT)} & \textbf{(nT)} & \textbf{(nT)} \\ \hline
1002.02 & 0.0  & 5.0  & 5.0  \\
1002.03 & -7.7 & 10.0 & 12.6 \\
1002.04 & -7.5 & 5.4  & 9.3  \\
1002.05 & -3.9 & 9.0  & 9.8  \\
1002.06 & 2.8  & 4.7  & 5.5  \\
1002.07 & 7.4  & 3.8  & 8.3  \\
1002.08 & 10.9 & 3.4  & 11.4 \\
1002.09 & 9.8  & 7.5  & 12.3 \\
1002.10 & 11.6 & 3.2  & 12.1 \\
1002.11 & 12.2 & 10.2 & 15.9 \\
1002.13 & 11.7 & 2.6  & 12.0 \\
1002.15 & 24.1 & 4.8  & 24.6 \\
1002.18 & 21.4 & 5.1  & 22.0 \\
1002.19 & 12.9 & 6.4  & 14.4 \\
1002.20 & 5.5  & 7.9  & 9.6  \\ \hline
\textbf{Avg} & \textbf{10.0} & \textbf{5.9} & \textbf{12.3} \\ \hline\hline
\end{tabular}
\end{table}

This appendix provides the comprehensive, line-by-line performance metrics across the fifteen evaluation flight trajectories of the DAF-MIT MagNav dataset. Table~\ref{tb_tl}, Table~\ref{tb_imp}, and Table~\ref{tb_kff3} present the detailed estimation bias, error standard deviation after removing the bias (Err. stdv), and Root Mean Square Error (RMSE) for the classical TL approach, the proposed augmented (AUG) calibration parameter estimation model, and the final KFF scheme, respectively. 

As demonstrated in Table~\ref{tb_tl}, the classical TL approach exhibits severe performance degradation on individual sensors, particularly \texttt{mag\_3} and \texttt{mag\_4}, yielding very large average RMSE value of 356.9~nT. This degradation is primarily driven by very large DC biases, which are eliminated by the band-pass filter during the calibration parameter estimation. These omitted portions are consequently impossible to capture during real-time compensation.

Table~\ref{tb_imp} confirms that the Stage 1 augmented linearized calibration parameter estimation successfully removes the bulk of these unmodeled sensor biases. It drastically reduces the total average RMSE across all sensors down from 359.2~nT to 58.0~nT.

Table~\ref{tb_kff3} presents the final multi-sensor joint estimation performance. Crucially, even though \texttt{mag\_3} and \texttt{mag\_4} remain relatively noisy, the KFF framework robustly leverages the clean characteristics of \texttt{mag\_5} (RMSE of 11.6~nT). The resulting multi-sensor fused average RMSE converges down to a highly accurate 12.3~nT. This demonstrates the exceptional structural robustness of the map-less state estimator, confirming its capability to prevent highly corrupted individual channels from degrading the global tracking performance.


\begin{thebibliography}{25}

\bibitem{Alken2021}
P. Alken et al.\\
International Geomagnetic Reference Field: the Thirteenth Generation\\
Earth, Planets and Space, 73, 49 (Feb. 2021).

\bibitem{BarShalom2001} 
Y. Bar-Shalom, X. R. Li and T. Kirubarajan\\ 
Estimation with Applications to Tracking and Navigation: Theory, Algorithms and Software.\\
New York, NY, USA: Wiley, 2001. 

\bibitem{BarShalom2011} 
Y. Bar-Shalom, P. K. Willett and X. Tian\\ 
Tracking and Data Fusion: A Handbook of Algorithms.\\
Storrs, CT, USA: YBS Publishing, 2011.

\bibitem{Bonifaz2020}
J. D. Bonifaz\\
Magnetic Navigation Using Online Calibration: Filter Analysis.\\
M.S. thesis,
Air Force Institute of Technology,
Wright-Patterson AFB, OH, USA, 2020.

\bibitem{Canciani2020}
A. J. Canciani and C. J. Brennan\\
An Analysis of the Benefits and Difficulties of Aerial Magnetic Vector Navigation.\\
IEEE Transactions on Aerospace and Electronic Systems, 56, 6 (Dec. 2020), 4161--4176.

\bibitem{Canciani2022} 
A. J. Canciani\\
Magnetic Navigation on an F-16 Aircraft Using Online Calibration.\\
IEEE Transactions on Aerospace and Electronic Systems, 58, 1 (Feb. 2022), 420--434. 

\bibitem{Dou2018}
Z. Dou, K. Ren, Q. Han and X. Niu\\
A Novel Real-Time Aeromagnetic Compensation Method Based on RLSQ.\\
In Proceedings of Tenth International Conference on Intelligent Information Hiding and Multimedia Signal Processing, 2014, 243--246.

\bibitem{Feng2022}
Y. Feng, Q. Zhang, Y. Zheng, X. Qu, F. Wu, and G. Fang\\
An Improved Aeromagnetic Compensation Method Robust to Geomagnetic Gradient.\\
Applied Sciences, 12, 3 (2022), 1490.

\bibitem{Ge2021} 
J. Ge, W. Wang, H. Dong, H. Liu, X. Zhang, W. Luo, H. Zhang, and J. Zhu\\
Modeling and reduction of the gradient effect of a magnetic sensor for an aeromagnetic compensator.\\
Review of Scientific Instruments, 92, 2 (2021), 024502. 

\bibitem{Gnadt2022}
A. R. Gnadt, A. B. Wollaber, and A. P. Nielsen\\
Derivation and Extensions of the Tolles-Lawson Model for Aeromagnetic Compensation.\\
arXiv:2212.09899, 2022. 

\bibitem{Gnadt2022t}
A. R. Gnadt\\
Advanced Aeromagnetic Compensation Models for Airborne Magnetic Anomaly Navigation.\\
Ph.D. dissertation,
Massachusetts Institute of Technology,
Cambridge, MA, USA, 2022.

\bibitem{Gnadt2023}
A. R. Gnadt \textit{et al.}\\
DAF-MIT AIA Open Flight Data for Magnetic Navigation Research.\\
Zenodo, https://zenodo.org/records/12723700, 2023.

\bibitem{Hager2026} 
A. Hager, S. Nebendahl, A. Klushyn, J. Krauser, T. H. Bryne, and T. A. Johansen\\
Airborne Magnetic Anomaly Navigation with Neural-Network-Augmented Online Calibration.\\
arXiv:2603.08265, 2026. 

\bibitem{Hager2026sub} 
A. Hager, T. H. Bryne, and M. Juki\'c\\ 
Scalar and Vector Airborne Platform Calibration Using Quantum and Classical Magnetometers and Inertial Sensors.\\
submitted to IEEE Transactions on Aerospace and Electronic Systems, 2026.

\bibitem{Han2017} 
Q. Han, Z. Dou, X. Tong, X. Peng, and H. Guo\\
A Modified Tolles--Lawson Model Robust to the Errors of the Three-Axis Strapdown Magnetometer.\\
IEEE Geoscience and Remote Sensing Letters, 14, 3 (Mar. 2017), 334--338.

\bibitem{Hezel2020}
M. A. Hezel\\
Improving Aeromagnetic Calibration Using Artificial Neural Networks.\\
Ph.D. dissertation,
Air Force Institute of Technology,
Wright-Patterson AFB, OH, USA, 2020.

\bibitem{Jiang2025}
Z. Jiang, T. Zhao, M. Wang, J. Zhou, Z. Deng and X. Lin\\
An Aeromagnetic Compensation Method Based on Differentiable Architecture Search-Guided Physics-Informed Neural Network.\\
IEEE Geoscience and Remote Sensing Letters, 22 (2025), 1--5.

\bibitem{Leliak1961} 
P. Leliak\\
Identification and Evaluation of Magnetic-Field Sources of Magnetic Airborne Detector Equipped Aircraft.\\
IRE Transactions on Aeronautical and Navigational Electronics, ANE-8 (1961), 95--105. 

\bibitem{Liu2025}
X. Liu, J. Liu, J. Zhang, W. Zhu, Q. Zhang, and G. Fang\\
INS-Based Aeromagnetic Compensation: Sliding Window Optimization for Enhanced Accuracy.\\ 
IEEE Access, 13 (2025), 138589--138600. 

\bibitem{MagNavjl2022}
USAF-MIT Artificial Intelligence Accelerator\\
MagNav.jl: airborne Magnetic anomaly Navigation.\\
https://github.com/MIT-AI-Accelerator/MagNav.jl, 2022.

\bibitem{Moradi2024}
M. Moradi, Z.-M. Zhai, A. Nielsen, and Y.-C. Lai\\
Random Forests for Detecting Weak Signals and Extracting Physical Information: A Case Study of Magnetic Navigation.\\
arXiv:2402.14131, 2024.

\bibitem{Tolles1950}
W. Tolles and J. Lawson\\
Magnetic Compensation of MAD Equipped Aircraft.\\
Airborne Instruments Laboratory, Mineola, NY, USA, Tech. Rep., 1950.

\bibitem{Wang2025} 
H. Wang and B. Zuo\\
An Aeromagnetic Compensation Algorithm Based on a Temporal Convolutional Network.\\
Applied Sciences, 15, 6 (2025), 3105.

\bibitem{Ye2024}
L. Ye, Z. Yu, Y. Zhang, C. Chi, P. Cheng, and J. Chen\\
An Aeromagnetic Compensation Strategy for Large UAVs.\\
Sensors, 24, 12 (2024), 3775.



\end{thebibliography}
\end{document}